\documentclass[aps,twocolumn,tightenlines]{revtex4}
\usepackage[dvipdf]{graphicx}
\usepackage{color}

\begin{document}
\title{Test of the He-McKellar-Wilkens topological phase by atom interferometry.\\
Part II: the experiment and its results.}

\author{S.~Lepoutre, J.~Gillot, A.~Gauguet, M.~B\"uchner, and J.~Vigu\'e}
\address{ Laboratoire Collisions Agr\'egats R\'eactivit\'e -IRSAMC
\\Universit\'e de Toulouse-UPS and CNRS UMR 5589
 118, Route de Narbonne 31062 Toulouse Cedex, France
\\ e-mail:~{\tt jacques.vigue@irsamc.ups-tlse.fr}}

\date{\today}

\begin{abstract}

In this paper, we describe an experimental test of the He-McKellar-Wilkens (HMW) topological phase with our lithium atom interferometer. The expected value of the  HMW phase shift in our experiment is small and its measurement was difficult because of stray phase shifts due to small experimental defects. We start by describing our setup and we characterize the effects of the electric and magnetic fields needed to observe the HMW effect. Then, we develop a model of our interferometer signals including all the defects we have identified. After various tests of this model, we use it to suppress the largest part of the stray phase shifts. We thus obtain a series of measurements of the HMW phase: the results are $31$\% larger than expected and this discrepancy is probably due to some limitations of our model.

\end{abstract}
\maketitle
\bigskip


\section{Introduction}
\label{intro}

The topological He-McKellar-Wilkens (HMW) phase introduced in 1993 by X.G. He and B.H.J. McKellar \cite{HePRA93} and in 1994
by M.~Wilkens \cite{WilkensPRL94} was never tested since its theoretical discovery. We have recently published such a test
\cite{LepoutrePRL12} using our lithium atom interferometer. In a companion paper \cite{LepoutreXXX} quoted here as HMWI,
we have recalled the theory of this topological phase and its relations with the Aharonov-Bohm \cite{AharonovPR59} and
Aharonov-Casher phases \cite{AharonovPRL84}. We have also discussed the effects of phase dispersion on interferometer signals and we have considered in detail the phase shifts induced by electric and magnetic fields, namely the dynamical phase shifts due to the Stark and Zeeman Hamiltonian and the topological phase shift due to the Aharonov-Casher effect. The present paper is devoted to a
detailed presentation of the experiment, of its results and of the analysis of the stray effects which have complicated the test of the HMW phase.

In the following sections, we first describe the experiment, the data recording procedure and the interferometer signals (section \ref{expset}). Then, we  discuss the effects of the electric fields (section \ref{elec}) and of the magnetic fields (section \ref{mag}). Section \ref{data} presents the data set for the HMW phase measurement and the raw results. The model describing the stray effects due to phase shift dispersion, introduced in HMWI and developed in the appendix (section \ref{App}), is tested thanks to numerous and sensitive measurements of the fringe phase and visibility (section \ref{test}). Thanks to this model, we have been able to reject most of the stray effects and to measure the HMW phase, as detailed in section \ref{HMWm}. A conclusion (section \ref{Conc}) summarizes what we have learnt from this experiment, recalls the open questions (in particular a phase shift presently not understood) and discusses how to improve this experiment.

\section{The experiment: the setup and the data recording procedure}
\label{expset}

In this part, we briefly describe our atom interferometer and, with greater details, the interaction region used to observe the HMW effect. We also describe the compensator coil used to produce a magnetic field gradient opposite to the one due to the HMW interaction region. Finally, we explain our data recording procedure which rejects the interferometer
phase drifts.

\subsection{Our atom interferometer}
\label{expset1}

%%%%%%%%%%%%%%%%%%%%%%%%%%%%%%%%%%%%%%%%%%%%%%%%%%%%%
\begin{figure}[t]
\begin{center}
\includegraphics[width= 8 cm]{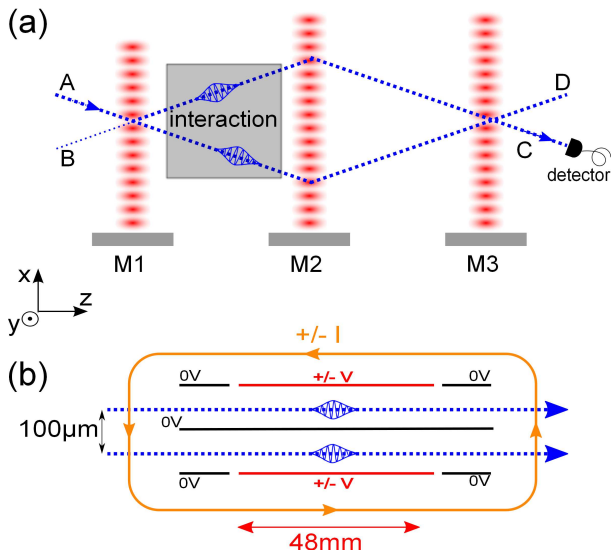} \caption{(color online)
Schematic top views (not to scale) of our atom interferometer (panel a) and the interaction region for the observation of the HMW phase (panel b). Our atom interferometer is a Mach-Zehnder interferometer, with two entrances A and B (only A is used) and two exits C and D (C is detected). An atomic beam (dotted blue lines) entering by A is diffracted by three quasi-resonant laser standing waves produced by the mirrors M$_i$. The largest distance between interferometer arms, about $100$ $\mu$m, occurs just before the second laser standing wave, where we introduce the interaction region. The opposite electric fields necessary for the observation of the HMW phase are horizontal. They are produced by two plane capacitors (high voltage electrodes in red; grounded electrodes in black). The septum is a thin common electrode located between the two interferometer arms represented by dotted blue lines. Two rectangular coils (represented by the brown rectangle) produce the vertical magnetic field.
\label{fig1}}
\end{center}
\end{figure}

Our separated arm atom interferometer (see figure \ref{fig1}), has been previously described \cite{MiffreEPJD05,MiffrePhD,LepoutreEPJD11}. Here, we present only its main features and some recent improvements. The atomic source is a supersonic beam of lithium seeded in argon,
with a mean velocity $v_m \approx 1065$ m/s. Once collimated by two $18$ $\mu$m wide slits, this beam has a transverse velocity distribution with a width comparable to the recoil velocity of lithium,  $v_r \approx 9$ cm/s. This beam is then
diffracted by three quasi-resonant laser standing waves in the Bragg regime: the present experiment uses first order diffraction, with only two diffracted beams of orders $0$ and $+1$ (or $-1$). We thus get a Mach-Zehnder atom interferometer with two output beams carrying complementary interference signals. One output beam is selected by a slit and its intensity $I$, measured by a surface ionization detector, is given by:

\begin{equation}
\label{a0} I=I_0 \left[1 + \mathcal{V}\cos\left(\varphi_d+\varphi_p\right)\right]
\end{equation}

\noindent $I_0$ is the mean intensity and $\mathcal{V}$ the fringe visibility. The phase is the sum of the phase $\varphi_p$ due to perturbations and the phase  $\varphi_d $ due to the diffraction process: $\varphi_d =2k_L(x_1-2x_2+x_3)$, where $k_L$ is the laser wavevector and $x_i$ the $x$-position of mirror M$_i$.  The choice of the laser frequency, at about $2$ GHz on the blue side of the $^2$S$_{1/2}$ - $^2$P$_{3/2}$ transition of $^7$Li, and the $92.5$\% natural abundance of $^7$Li ensure that the interferometer signal is almost purely due to this isotope \cite{MiffreEPJD05,JacqueyEPL07}. To record interference fringes, we sweep the phase $\varphi_d $ by varying $x_3$ with a piezoelectric actuator. We measure the variations of $x_3$ with a Michelson interferometer
\cite{LepoutreEPJD11}. Intense signals, with a mean intensity $I_0 \approx 60000$ atoms/s and a high fringe visibility $\mathcal{V} \approx 70$\% provide a large phase sensitivity, with a practically achieved value $\Delta\varphi_{min} \approx 25$ mrad$/\sqrt{ \mbox{Hz}}$.

\subsection{The HMW interaction region}
\label{expset2}

A HMW phase is induced when an atom propagates in crossed electric and magnetic fields, both transverse to the atom velocity. Our experimental arrangement is inspired by the ideas of H. Wei \textit{et al.} \cite{WeiPRL95}, the electric fields are horizontal, in the interferometer plane, and opposite on the two interferometer arms, while the common magnetic field is vertical and as homogeneous as possible.

\begin{figure}
\begin{center}
\includegraphics[width= 8 cm]{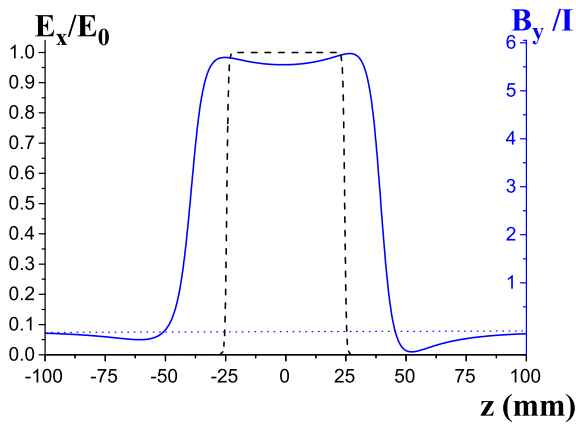}
\caption{(color online) Calculated components $E_x$ (dashed line) and $B_y$ (full line) as a function of $z$ in the interaction region. For the electric field, the plotted quantity is $E_x/E_0$, where $E_0 = V/h$ is the field of a infinite plane capacitor, with a spacing $h$ and an applied voltage $V$. For the magnetic field, the plotted quantity $B_y/I$, where $I$ is the coil current, is in units of $10^{-4}$ T/A. \label{fig2}}
\end{center}
\end{figure}

The electric fields are produced by a double capacitor with a septum \cite{Ekstrom95} located between the interferometer
arms (see figure \ref{fig1}). Each of the two capacitors is similar to the one we used for the measurement of the electric
polarizability of lithium \cite{MiffreEPJD06}. Outer electrodes are made of polished $5$ mm-thick glass plates with evaporated aluminium electrodes. A central high-voltage electrode of length $2a \approx 48$ mm is separated from two $5$ mm-long grounded guard electrodes by $1$ mm-wide gaps: these gaps withstand a voltage larger than $1$ kV. The septum is a $30$ $\mu$m thick aluminium foil. The capacitors are assembled by gluing together the electrodes and the glass spacers (thickness $h\approx 1.10$ mm) with Epotex 301 glue (Epoxy Technologies). The septum must remain well stretched, even if the capacitor temperature varies. With some advice given by A. Cronin, we have acquired the know-how to glue a pre-stretched septum on the electrode-spacer assembly heated near $65^{\circ}$C and, due to differential thermal expansion, the septum is fully stretched when the assembly has cooled down \cite{LepoutrePhD11}. The capacitors are as symmetric as possible and they are powered by slightly different voltages issued from the same power supply, with an adjustable voltage ratio thanks to potentiometers. This arrangement minimizes Stark phase noise due to voltage fluctuations of the power supply. Figure \ref{fig2} presents the calculated $z$-variation of the electric field $E_x$-component, which is relevant for the HMW phase. $E_x$ is calculated at the septum surface whereas the atom-septum mean distance is near $40$ $\mu$m but the associated correction is very small \cite{MiffreEPJD06}.

The capacitors assembly is inserted in a brass support on which we have coiled $1.5$ mm-diameter enameled copper wires to produce the vertical magnetic field needed for the HMW phase. We use two rectangular coils, located below and above the interferometer plane, each coil being made of $2$ layers and each layer of $7$ windings, glued together and to the brass support with a high thermal conductivity glue (Stycast 2850 FT). A $2$-mm internal diameter copper pipe is also glued on the brass support at mid-distance between the two coils and with a $6$ cm$^3$/s flow of tap water (a low flow rate chosen to minimize vibrations),
the temperature rise is about $0.5$ K/W. Usually, we apply a current $I$ in the coils $50$\% of the time so that the maximum current $I = 25$ A induces a $20$ W mean Joule power and a temperature rise near $10$ K. In figure \ref{fig2}, we have plotted the calculated $z$-variation of the magnetic field $B_y$-component which is the one relevant for the HMW phase. As discussed in
HMWI, a Zeeman phase shift appears if the magnetic field modulus $B$ is different on the two interferometer arms, and we have minimized this difference by careful coiling and design of the connection wires geometry.

%%%%%%%%%%%%%%%%%%%%%%%%%%%%%%%%%%%%%%%%%%%%%%%%%%%%%
%%%%%%%%%%%%%%%%%%%%%%%%%%%%%%%%%%%%%%%%%%%%%%%%%%%%%%:

The HMW interaction region is placed just ahead the second laser standing wave, where the distance between the center of the interferometer arms is largest, close to $100$ $\mu$m. In order not to induce vibrations of the standing wave mirrors, the interaction region is suspended from the top of the vacuum chamber.  Initial adjustments of the rotation around the horizontal $z$-axis and the vertical $y$-axis are performed with optical methods. Rotation around $y$-axis as well as translation in the $x$-direction can be operated under vacuum, and the ultimate tunings are done with the atom interferometer running. After optimization of the interferometer signal, the mean intensity $I_0$ and the fringe visibility $\mathcal{V}$ are not modified by the presence of the septum between the two arms.

The magnetic field produced by the HMW coil was measured with a 3D Hall probe and compared to the field calculations, showing
a good agreement. Concerning the electric field, calibration measurements using the atom interferometer (described in part
\ref{elec1}) yield an accurate knowledge of each capacitor geometry needed for electric field calculations. With the electric and
magnetic field components $E_x$ and $B_y$ as a function of $z$, we can calculate the integral $\int E_x B_y dz$ and thanks to the very accurate knowledge of the electric polarizability of lithium atom \cite{MiffreEPJD06,PuchalskiPRA12}, we can predict the
slope of the HMW phase as a function of $VI$ product:

\begin{equation}
\label{a01} \varphi_{HMW}  \left(V, I \right)/(VI) = - (1.28 \pm 0.03) \times 10^{-6} \mbox{ rad/VA}
\end{equation}

\noindent where the error bar is due to the uncertainty on the geometrical parameters of the capacitors and of the HMW coil.

\subsection{Compensator coil}
\label{expset3}

In spite of our efforts, the magnetic field of the HMW coil is slightly different on the two interferometer arms, with a mean
relative difference $\left|\Delta B \right|/ B \approx 10^{-4}$. This difference is most probably due to a bad centering of the septum in the HMW coil, with a distance between the coil symmetry plane and the septum of the order of $250$ $\mu$m. For $I=25$ A, the induced Zeeman phase shift is equal to  $\varphi_{Z}(F,m_F)\approx \pm 11$ rad for the $F=2, m_F= \pm 2$ sublevels. We compensate these phase shifts thanks to a supplementary coil producing an opposite magnetic field gradient along the $x$-axis. This so-called compensator coil is made of $9$ turns of copper wire on a $30$ mm-diameter aluminium cylinder. It is located at mid-distance between the first and second laser standing waves, with a mean distance between the compensator coil and the interferometer arms near $10$ mm. This coil is cooled by conduction through its support and temperature rise limits its current $I_C$ to $5$ A, if applied only $50$\% of the time. Then, the magnetic field seen by the atoms is below $2\times 10^{-3}$ T, a range for which Zeeman effect is linear.

\subsection{Data recording and signals}
\label{expset4}

%%%%%%%%%%%%%%%%%%%%%%%%%%%%%%%%%%%%%%%%%%%%%%%%%%%%%
\begin{figure}
\begin{center}
\includegraphics[width= 8 cm]{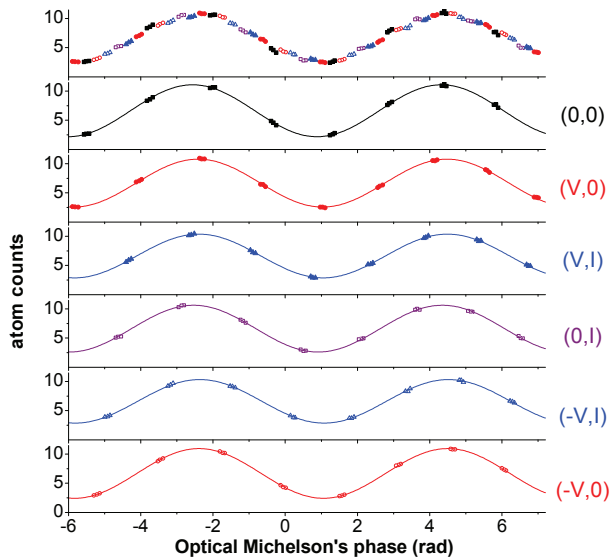} \caption{ (color online) Recorded data with 6 ($V,I$) configurations during a single fringe scan: the number of detected atoms per second (the unit is $10^4$ detected atoms/second and the counting period is $0.1$ second) is plotted as a function of the phase of the optical Michelson interferometer which directly maps the position $x_3$ of mirror M$_3$. The signals corresponding to different field configurations are plotted with different symbols : black squares for $(0,0)$, red bullets for $(V,0)$, blue full triangle for $(V,I)$, open violet squares for $(0,I)$, open blue triangle for $(-V,I)$ and open red circles for $(-V,0)$). The top graph represents the signal as it is recorded and the signal corresponding to each one of the 6 ($V,I$) configurations is plotted separately below with its best fit. \label{fig3}}
\end{center}
\end{figure}
%%%%%%%%%%%%%%%%%%%%%%%%%%%%%%%%%%%%%%%%%%%%%%%%%%%%%%

In our previous experiments \cite{MiffreEPJD06,JacqueyEPL07,JacqueyPRL07,LepoutreEPJD11}, we deduced the effect of a perturbation by comparing fringe signals successively recorded with and without this perturbation. The phase measured in the absence of perturbation, which should be constant, drifts with time, typically by several tens of mrad over the few minutes needed for recording a high-quality fringe signal. These drifts are not linear in time and they are due to minute distortions of the rail supporting the standing wave mirrors. Their magnitude is due to the high sensitivity of the diffraction phase $\varphi_d$ to the mirror positions, with $d\varphi_d/ dx_i \approx + 20$ rad/$\mu$m  for M$_1$ or M$_3$ and $-40$ rad/$\mu$m for M$_2$. They were the main limitation of our phase shift measurements. For the present experiment, we have almost canceled the sensitivity to these drifts by applying several field configurations during each fringe recording: a field configuration is defined by the $(V,I)$
values, where $V$ is the capacitor mean voltage and $I$ the current in the HMW coil (this current is accompanied by a current $I_C$ in the compensator coil, as explained below). We have used either 4 configurations, $(0,0)$, $(V,0)$, $(V,I)$ and $(0,I)$ or 6 configurations, by adding $(-V,I)$ and ($-V,0)$ to this list. A typical fringe recording with $6$ configurations is shown in figure \ref{fig3}. A fit of the different fringe signal systems is made using equation (\ref{a0}), where the fringe systems of all configurations share the same value of the diffraction phase $\varphi_d$. We thus get the mean intensity $I_0(V,I)$, the fringe phase $\varphi(V,I)$ and the fringe visibility $\mathcal{V}(V,I)$ for each field configuration. In this way, we deduce the effects of the application of the electromagnetic field corresponding to each configuration. $I_0$ is independent of the field  configuration, but the visibility and the phase are both modified. We define a relative visibility and a fringe phase shift for each field configuration by:

\begin{eqnarray}
\label{a1}
\mathcal{V}_E(V) &=& \mathcal{V}(V,0)/\mathcal{V}(0,0) \nonumber \\
\varphi_E(V) &=& \varphi(V,0) - \varphi(0,0)\nonumber \\
\mathcal{V}_B(I) &=& \mathcal{V}(0,I)/\mathcal{V}(0,0) \nonumber \\
\varphi_B(I) &=& \varphi(0,I) - \varphi(0,0)\nonumber \\
\mathcal{V}_{EB}(V,I) &=& \frac{\mathcal{V}_{E+B}(V,I)}{\mathcal{V}_E(V) \mathcal{V}_B(I)} =\frac{\mathcal{V}(V,I)\mathcal{V}(0,0)}{\mathcal{V}(V,0) \mathcal{V}(0,I)} \nonumber \\
\varphi_{EB}(V,I) &=& \varphi_{E+B}(V,I)- \varphi_E(V)- \varphi_B(I) \nonumber \\
&=& \varphi(V,I) - \varphi(V,0)-\varphi(0,I)+\varphi(0,0)\nonumber \\
\end{eqnarray}
\noindent

\noindent $\mathcal{V}_{E}(V)$ and $\varphi_E(V)$ are the fringe relative visibility and phase shift with the electric field only. $\mathcal{V}_{B}(I)$ and $\varphi_B(I)$ are the fringe relative visibility and phase shift with the magnetic field only and
$\mathcal{V}_{E+B}(V,I)$ and $\varphi_{E+B}(V,I)$ are the fringe relative visibility and phase shift with the electric and magnetic fields applied simultaneously. A fringe scan such as shown in figure \ref{fig3} lasts about $20$ s, a duration sufficiently small to ensure quasi-linearity of the interferometer phase drift with time (an exactly linear phase drift only alters $\varphi_d$ and leaves the results of eqs. (\ref{a1}) unchanged). The error bars are about $2$\% on the relative visibility and $30$ mrad on the induced phase shifts. We repeat about $100$ successive fringe scans, taking care that the fringe scan period and the field configuration period are not commensurate, in order to avoid any possible bias in the fits. The error bars on the averages of such a scan series are near $0.2$\% for the relative visibility and near $3$ mrad for the phase shifts, small enough to detect fine
perturbations of the interference fringe signals and to understand systematic effects.

\section{Effects of the electric fields on the fringe phase and visibility}
\label{elec}

\subsection{Experimental study of polarizability phase shifts}
\label{elec1}

\begin{figure}
\begin{center}
\includegraphics[width = 8 cm]{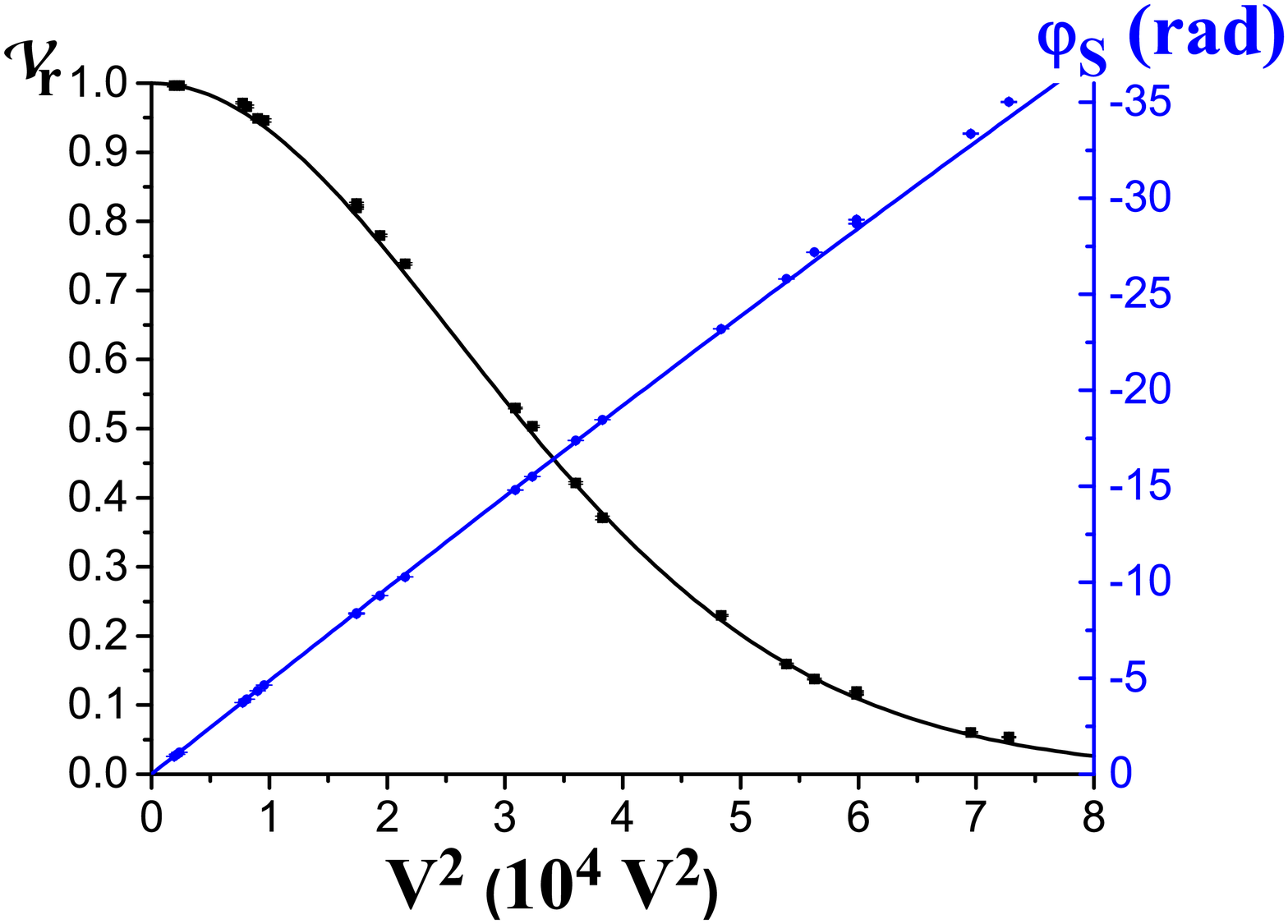} \caption{ (color online) The relative visibility $\mathcal{V}_r$ (left scale, black squares and line) and the fringe phase shift $\varphi_{S}$ (rad) (right scale, blue bullets and line) are plotted as a function of $V^2$, where $V$ is the voltage applied to one capacitor only. The points are experimental and the curves are their best fits.
\label{fig4}}
\end{center}
\end{figure}

During calibration measurements, we applied a voltage $V$ to one capacitor only, the other one being grounded. Figure \ref{fig4} presents typical results for the fringe visibility $\mathcal{V}_r = \mathcal{V}(V)/\mathcal{V}(0)$ and the induced phase shift $\varphi_S (V)$ as a function of $V^2$. These measurements were fitted using equations (19-20) and (31) of HMWI, which yields the
value of the parallel speed ratio $S_{\|}= 9.25 \pm 0.08$ and the values of the Stark phase shifts induced by each capacitor for the mean atom velocity: $\varphi_u/V^2 = (-4.830 \pm 0.005)$ rad/V$^2$ and $\varphi_l/V^2 = (4.760 \pm 0.007)$ rad/V$^2$. Using the very accurate theoretical value \cite{PuchalskiPRA12} of lithium atom electric polarizability $\alpha$, we may deduce the geometrical parameter $L_{eff}/h^2$ for both capacitors ($L_{eff}$ is the capacitor effective length and $h$ the plate spacing
\cite{MiffreEPJD06}). The effective length is the same for both capacitors with a good accuracy, $L_{eff}\approx 48\pm 0.5 $ mm, so that these experiments provide measurements of the mean values of the capacitor spacings $h_u = 1.101\pm 0.006$ mm and $h_l= 1.109\pm 0.006$ mm.

%%%%%%%%%%%%%%%%%%%%%%%%%%%%%%%%%%%%%%%%%%%%%%%%%%%%%
%%%%%%%%%%%%%%%%%%%%%%%%%%%%%%%%%%%%%%%%%%%%%%%%%%%%%%

\subsection{Experiments with both electric fields on: phase measurements}
\label{elec2}

%%%%%%%%%%%%%%%%%%%%%%%%%%%%%%%%%%%%%%%%%%%%%%%%%%%%%
\begin{figure}
\begin{center}
\includegraphics[width = 8 cm]{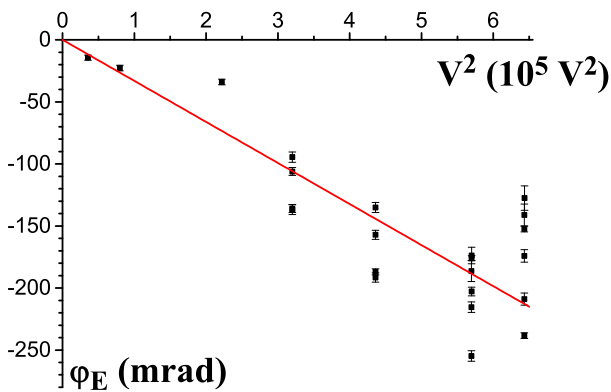} \caption{(color online) Fringe phase shifts induced by electric fields applied to both capacitors: the measured phase shift $\varphi_E(V)$ is plotted as a function of $V^2$ where $V$ is the mean of the voltages $V_u$ and $V_l$ applied to the two capacitors. The points are experimental and the straight line is the best fit. \label{fig5}}
\end{center}
\end{figure}
%%%%%%%%%%%%%%%%%%%%%%%%%%%%%%%%%%%%%%%%%%%%%%%%%%%%%%

When we apply electric fields to both capacitors, with the voltage ratio tuned to cancel the Stark phase shift $\varphi_E(V)$, we
observe a residual phase shift due to imperfect tuning: $\varphi_E(V)$ is small and approximately proportional to $V^2$, but with large fluctuations of the measured value (see figure \ref{fig5}). We have observed that $\varphi_E(V)$ drifts with time when the interaction region temperature varies: this behavior can be explained by a delay of the expansion of one capacitor with respect
to the other one, delay due to the low thermal conductivity of glass. For $V=800$ V, the Stark phase induced on each
interferometer arm can reach $\varphi_{S,u} \approx \varphi_{S,l} \approx 307$ rad. Then, a typical deviation for $\varphi_E(V)$ of $0.05$ rad from its mean value corresponds to a $1.7 \times 10^{-4}$ relative variation of the geometrical parameter
$L_{eff}/h^2$ of one capacitor with respect to the other one. This variation is somewhat larger than expected for a simple thermal expansion effect with a temperature variation smaller than $10$K. The conclusion is that, because of dispersion and drift,
the residual Stark phase shift $\varphi_E(V)$ does not carry much useful information.

%%%%%%%%%%%%%%%%%%%%%%%%%%%%%%%%%%%%%%%%%%%%%%%%%%%%%
\begin{figure}
\begin{center}
\includegraphics[width = 8 cm]{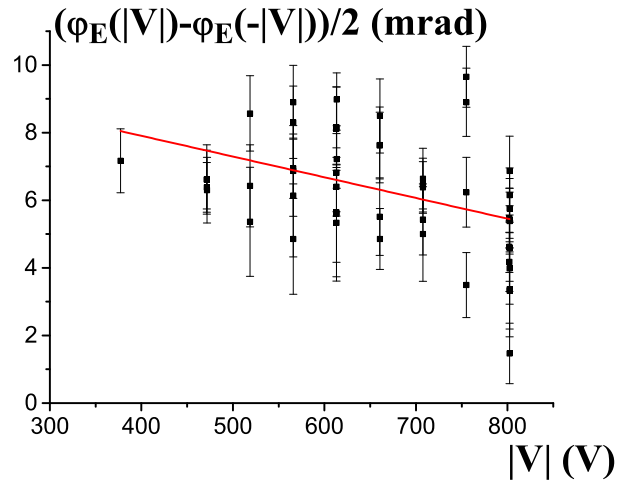} \caption{(color online) Fringe phase shifts induced by electric fields applied to both capacitors : the difference quantity $\left(\varphi_E(\left|V\right|)-\varphi_E(-\left|V\right|)\right)/2$ is plotted as a function of $\left|V\right|$. The points are experimental and the straight line is the best linear fit.
\label{fig6}}
\end{center}
\end{figure}
%%%%%%%%%%%%%%%%%%%%%%%%%%%%%%%%%%%%%%%%%%%%%%%%%%%%%%

In the experiments with 6 field configurations, we can measure the difference of the Stark phase shifts for opposite $V$ values, with an error bar close to $1$ mrad.  Figure \ref{fig6} plots the quantity $\left(\varphi_E(\left|V\right|)
-\varphi_E(-\left|V\right|)\right)/2$ as a function of $\left|V\right|$. Eq. (58) of HMWI predicts that the only $V$-odd
term in $\left<\varphi_S\right>$ is the contact potential phase $\left<\varphi_{Sc}\right> $ so that:

\begin{equation}
\label{ex0} \left(\varphi_E(\left|V\right|) -\varphi_E(-\left|V\right|)\right)/2 = 2 \varphi_{0} \frac{\left<\bar{V}_{c,u}\right> -\left<\bar{V}_{c,l}\right>}{V}
\end{equation}

\noindent  We have fitted the measured values of $\left(\varphi_E(\left|V\right|) -\varphi_E(-\left|V\right|)\right)/2$ by a function $a+b\left|V\right|$. The fitted $a$-value, $a= 10 \pm 1$ mrad is not explained by eq. (\ref{ex0}) but its  presence  might be due to the use of different power supplies, one per polarity, to produce opposite voltages. The fitted slope $b= (-6 \pm 2) \times 10^{-3}$ mrad/V can be explained by a difference of the mean contact potentials $\left(\left<\bar{V}_{c,u}\right>
-\left<\bar{V}_{c,l}\right>\right) = 6\pm 2$ mV: we may conclude that contact potentials play a very minor role in our experiment and this idea will be supported by further results.

\subsection{Experiments with both electric fields on: visibility measurements}
\label{elec3}

%%%%%%%%%%%%%%%%%%%%%%%%%%%%%%%%%%%%%%%%%%%%%%%%%%%%%
\begin{figure}
\begin{center}
\includegraphics[width = 8 cm]{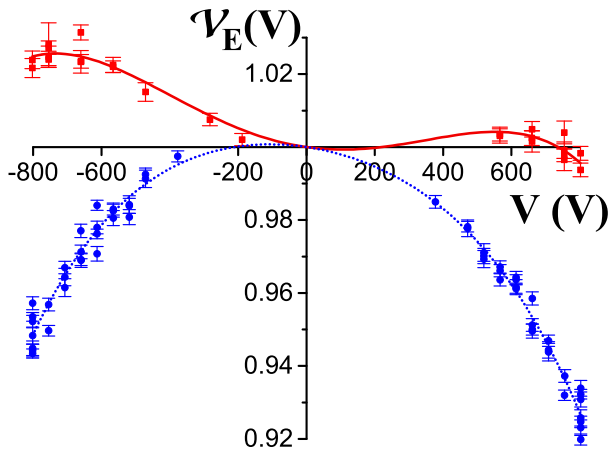} \caption{(color online) Fringe relative visibility $\mathcal{V}_E(V)$ as a function of the applied mean voltage $V$: the points are experimental, with different symbols (red squares and blue bullets) for two different alignments of the atom interferometer. The data points are fitted by eq. (\ref{ex1}) (red full line and blue dotted line). \label{fig7}}
\end{center}
\end{figure}
%%%%%%%%%%%%%%%%%%%%%%%%%%%%%%%%%%%%%%%%%%%%%%%%%%%%%%

We now discuss the measurements of the relative visibility $\mathcal{V}_E(V)$. The residual Stark phase shift
$\varphi_E$ is sufficiently small to neglect any effect of the velocity dispersion on the visibility. The relative visibility showed highly dependent of the standing wave mirrors alignment. Therefore the collected data for $\mathcal{V}_E(V)$ was partitioned into 7 different sets: within a given set, this interferometer alignment is identical for all the data points. Fig. \ref{fig7} presents two of these data sets. All the sets exhibit different behaviors, but are well independently fitted using the following equation:

\begin{eqnarray}
\label{ex1}
\mathcal{V}_E(V)= 1 -\sum _{i=1,4} k_{Vi} V^i\nonumber \\
\end{eqnarray}

\noindent As illustrated by fig. \ref{fig7}, the relative visibility can become larger than $1$, a result apparently surprising. This happens when phase dispersions which exist when no interaction is applied, are partially canceled by the phase dispersion due to the application of the electric fields. The pre-existing phase dispersions originate from the Zeeman phase shifts due to the inhomogeneity of the laboratory magnetic field when $I= I_C = 0$ and from the diffraction phase which presents a spatial dispersion because of an imperfect alignment of the laser standing wave mirrors M$_i$. The application of the electric fields induces an Aharonov-Casher phase shift and a Stark phase shift: the Aharonov-Casher phase shift is dispersed because of its dependence with the $F,m_F$ sublevel and the Stark phase shift is dispersed because of capacitors defects. We must describe all these effects, in order to explain the behavior of $\mathcal{V}_E(V)$. Assuming a balanced hyperfine population i.e. $\chi=0$ ($\chi$ is defined in appendix A of HMWI), we use equations (23), (42) and (46-47) of HMWI to evaluate $\mathcal{V}(0,0)$:

\begin{eqnarray}
\label{ex2} \frac{\mathcal{V}(0,0)}{\mathcal{V}_{0}} &=& \left[1 -
\frac{\left< \left(\delta \varphi_d\right)^2\right>}{2}\right]
\frac{\mathcal{V}_{B0}}{\mathcal{V}_0} \nonumber \\
\mbox{with  } \frac{\mathcal{V}_{B0}}{\mathcal{V}_0} &=& \frac{1 +
\cos\left(J_0\right) + 2 \cos\left(J_0/2\right)}{4}
\end{eqnarray}

\noindent $\delta \varphi_d (y)$ is the dispersion of the diffraction phase and $\mathcal{V}_{B0}$ is the visibility modified
by the inhomogeneity of the laboratory magnetic field. Because of these two effects, the observed visibility $\mathcal{V}(0,0)$ is smaller than its optimum value $\mathcal{V}_0$. With electric fields on both arms, the Stark phase shift $\varphi_S (y)$ is a function of $y$, and the Aharonov-Casher phase shift $\varphi_{AC}(F,m_F)$ is given by $\varphi_{AC}(2,m_F) = -\varphi_{AC}(1,m_F) = (m_F/2) \varphi_{AC}(2,2)$ (this formula is valid because Zeeman effect is linear in the laboratory field). We deduce the fringe visibility $\mathcal{V}(V,0)$:

\begin{eqnarray}
\label{ex3} \frac{\mathcal{V}(V,0)}{\mathcal{V}_0}  &=&
\mathcal{V}_{B0}\left[1 - \frac{\left< \left(\delta \varphi_d+
\delta\varphi_S\right)^2\right>}{2}\right] \nonumber \\
&-& \frac{\varphi_{AC}(2,2)}{4} \left[\sin\left(J_0\right) +
\sin\left(\frac{J_0}{2}\right)\right]
\end{eqnarray}

\noindent The last term of the r.h.s. is a first-order Taylor expansion of the trigonometric functions of $\varphi_{AC}$, valid because $\left|\varphi_{AC} \right| \leq 70$ mrad in the present experiment. We get the relative visibility $\mathcal{V}_E(V)= \mathcal{V}(V,0)/ \mathcal{V}(0,0)$:

\begin{eqnarray}
\label{ex4} \mathcal{V}_E(V) &=&  1 - \frac{\left< \left(\delta
\varphi_S\right)^2+ 2 \delta
\varphi_d\delta\varphi_S\right>}{2}\nonumber \\
&& - \frac{\varphi_{AC}(2,2)}{4\mathcal{V}_{B0}}
\left[\sin\left(J_0\right) + \sin\left(\frac{J_0}{2}\right)\right]
\end{eqnarray}

\noindent The dispersion $\delta\varphi_S$ of the Stark phase shift is given by $\delta\varphi_S = \delta \varphi_{S,g}+\delta
\varphi_{S,c} $ with the geometrical defect term $\delta \varphi_{S,g} \propto V^2$ and the contact potential term $\delta
\varphi_{S,c} \propto V$. $\varphi_{AC}(2,2) \propto V$ while $\delta \varphi_d$ and $J_0$ are independent of $V$. We thus deduce the values of the $k_{Vi}$ coefficients:

\begin{eqnarray}
\label{ex5}
k_{V1} &=& \left< \delta \varphi_d\delta\varphi_{S,c}\right> + \frac{\varphi_{AC}(2,2)}{4\mathcal{V}_{B0}} \left[\sin\left(J_0\right) +  \sin\left(\frac{J_0}{2}\right)\right] \nonumber \\
k_{V2} &=&  \left< \left(\delta \varphi_{S,c}\right)^2/2\right> + \left<\delta \varphi_d \delta\varphi_{S,g}\right> \nonumber \\
k_{V3} &=& \left< \delta \varphi_{S,g}\delta \varphi_{S,c}\right>  \nonumber \\
k_{V4} &=& \left< \left(\delta \varphi_{S,g}\right)^2/2\right>
\end{eqnarray}

\noindent Discussed below is a comparison of eqs. (\ref{ex5}) with the results of fits of $\mathcal{V}_E (V)$ for
the 7 available data sets: at the same time, we test the validity of our description of experimental defects and we get some insights on the nature of the systematic effects.

All $k_{V4}$ values are positive and compatible with their mean,
$k_{V4} = (6.0 \pm 0.5)\times 10^{-14}$ V$^{-4}$. This is in
agreement with  eqs. (\ref{ex5}) which predicts that $k_{V4}$ is
positive and depends solely of the geometrical defects of the
capacitors and not of the interferometer alignment. From this result,
we may estimate the geometrical defects of the capacitors if we
assume that the spacing difference $\Delta h =h_u-h_l $ is the main
defect and that it varies linearly with $y$. We then find that
$\Delta h $ varies by about $1.4$ $\mu$m over the $y$-range sampled
by the atoms (about $2$ mm). This $\Delta h$ value appears to be quite small for capacitors
assembled by gluing parts together but, when $V = 800$ V, this small defect is sufficient to induce a total dispersion of the Stark phase shift along the atomic beam height equal to $0.8$ rad.

All $k_{V3}$  values (excepted one) are compatible with $0$, with a
very small mean value, $k_{V3}= (0.04 \pm 1.7)\times 10^{-12}$
V$^{-3}$, corresponding to $\left< \delta \varphi_{S,g}\delta
\varphi_{S,c}\right> < 10^{-6}$ rad$^2$ for $V=800$ V. The
dispersions $\delta \varphi_{S,g}$ and $\delta \varphi_{S,c}$ are not
correlated, in agreement with the idea that contact potentials
fluctuate on small scales and that geometrical defects are smooth
functions of $y$.

Each $k_{V2}$ value has a small error bar but $k_{V2}$ varies
strongly from one set of data to the next, covering the range from
$-5\times 10^{-8}$ to $+13\times 10^{-8}$ V$^{-2}$.
These large variations prove that the dominant contribution comes
from the interferometer alignment i.e. from the $\left<\delta \varphi_d
\delta\varphi_{S,g}\right>$ term. When $\delta \varphi_d$ and $\
\delta\varphi_{S,g}$ have opposite variations, the electric fields
increase the visibility, as observed in figure \ref{fig7}.

All $k_{V1}$ values are  compatible with their mean, $k_{V1}=
\left(1.40 \pm 0.07\right) \times 10^{-5}$ V$^{-1}$. The first term
$\left< \delta \varphi_d\delta\varphi_{S,c}\right>$, which involves
the contact potential term, is expected to be very small for the same
reasons which explain the weakness of $k_{V3}$ and, if this term was
not negligible, $k_{V1}$ should vary with the interferometer
alignment like $k_{V2}$. The second term, which is due to the
Aharonov-Casher phase shift in the laboratory magnetic field, must be
dominant. Assuming that the laboratory magnetic field $\mathbf{B}_0$
is constant over the capacitor length, and that the electric fields
are equal to $E_0= V/h$ on a length $L_{eff}\approx 48$ mm, we
estimate the Aharonov-Casher phase shift given in
eqs. (2) and (49) of HMWI:

\begin{eqnarray}
\label{ex6}
\varphi_{AC}(2,2) = \frac{2\mu_B E_0 L_{eff}}{\hbar c^2}\mathbf{y}\cdot \mathbf{u}_0
\end{eqnarray}

\noindent where $\mathbf{u}_0 =\mathbf{B}_0/B_0$ points in the
direction of $\mathbf{B}_0$. The measurements presented in the next
section give access to $J_0 \approx -0.61$ rad and to
$\mathcal{V}_{B0}\approx 0.93$. We thus deduce $\mathbf{y}\cdot
\mathbf{u}_0 \approx -0.7$ i.e. $\mathbf{B}_0$ points downward, at
about $45^{\circ}$ from the vertical, in agreement with direct
measurements of the local laboratory field.

\section{Effects of the magnetic field on the fringe phase and visibility}
\label{mag}

\subsection{Experiments with the compensator coil only}
\label{mag1}

\begin{figure}
\begin{center}
\includegraphics[width = 8 cm]{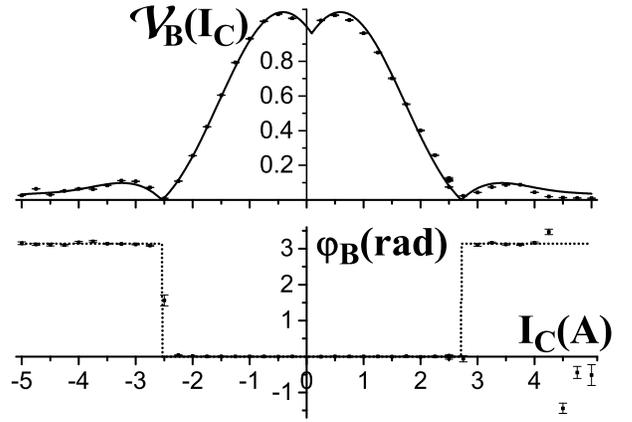} \caption{Relative visibility (upper panel) and phase shift (lower panel) as a function of the compensator current $I_C$ (A). The points are experimental and the curves represent their best fit, with $A_{J1,C} = 1.52 \pm 0.02$ rad/A, $I_{0,C} = 0.09 \pm 0.02$ A and $J_{0,C} = -0.63 \pm 0.03$ rad. The minor deviations which appear when $I_C> 4$ A are
probably due to the arbitrary assumption of balanced sublevel populations ($\chi=0$) for this particular example. \label{fig8}}
\end{center}
\end{figure}

We have measured the relative visibility $\mathcal{V}_B(I_C)$ and the phase shift $\varphi_B(I_C)$ of the interference fringes as a function of the compensator coil current $I_C$. The results are plotted in fig. \ref{fig8} with fits based on eqs. (42) and (45) of HMWI, with $J_1 = A_{J1,C} \left|I_{C}-I_{0,C}\right| + J_{0,C}$ and assuming balanced sublevel populations (see appendix A of HMWI).

%%%%%%%%%%%%%%%%%%%%%%%%%%%%%%%%%%%%%%%%%%%%%%%%%%%%%
%%%%%%%%%%%%%%%%%%%%%%%%%%%%%%%%%%%%%%%%%%%%%%%%%%%%%%

\subsection{Experiments with the HMW-coil only}
\label{mag2}

\begin{figure}
\begin{center}
\includegraphics[width = 8 cm]{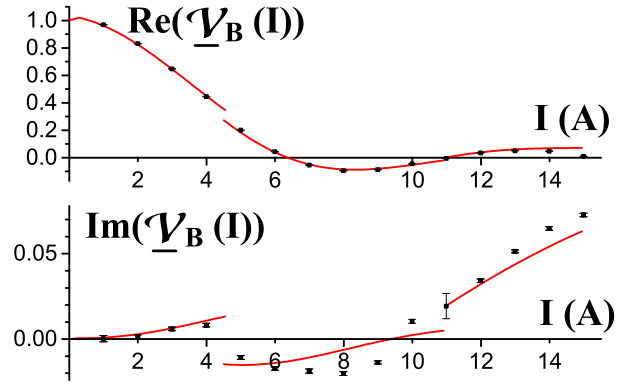} \caption{(color online) Real and imaginary part of the complex fringe visibility as a function of the HMW coil current $I$ (note the expanded scale for Im$(\mathcal{V}_B(I)$). The points are experimental and the curves are the results of  best fits, with 3 different $\chi$ values: $\chi= 0.077$ for $I= 0-4$ A; $\chi= -0.014$ for $I= 5-10$ A; $\chi= -0.062$ for $I= 11-15$ A. When these experiments were done, we had not understood that the laser frequency must be tightly controlled in order to keep $\chi$ very small and this explains why large $\chi$ values are observed. We get  $A_{J1} = -0.46 \pm 0.02 $ rad/A; $A_{J2} = \left(-110 \pm 8\right)\times
10^{-4}$ rad/A$^2$; $A_{J3}= \left(-20 \pm 4 \right)\times 10^{-5}$
rad/A$^3$; $I_{0} = 0.32 \pm 0.05$ A and $J_{0,I} = -0.55 \pm 0.13$ rad. \label{fig9}}
\end{center}
\end{figure}

The relative visibility $\mathcal{V}_B(I)$ and the phase shift
$\varphi_B(I)$ were measured as a function of the HMW coil current
$I$. Some of the results are plotted in fig. \ref{fig9}.
Equation (41) of HMWI gives $\varphi_{Z}(F,m_F)$ as a function of
$J_1$, $J_2$ and $J_3$ and we use  $J_1 =  A_{J1}
\left|I-I_{0} \right| + J_{0,I}$, $J_2= A_{J2} I^2$, $J_3= A_{J3}
\left|I\right|^3$ to fit the data.  The hyperfine
population unbalance parameter $\chi$ is also fitted, with a different
value for each data set corresponding to a slightly different laser frequency.

%%%%%%%%%%%%%%%%%%%%%%%%%%%%%%%%%%%%%%%%%%%%%%%%%%%%%
%%%%%%%%%%%%%%%%%%%%%%%%%%%%%%%%%%%%%%%%%%%%%%%%%%%%%%

\subsection{Experiments with both coils and global fit}
\label{mag3}

\begin{figure}
\begin{center}
\includegraphics[width = 8 cm]{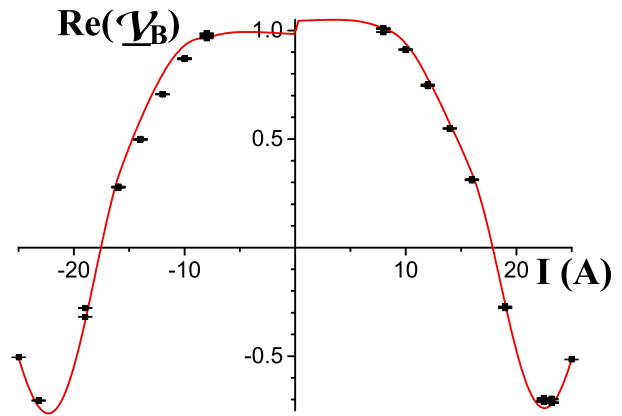} \caption{(color online) Real part of the complex relative visibility $Re\left(\underline{\mathcal{V}}_B\right)(I)$  plotted as a function of the HMW current $I$, the compensator current $I_C$ having the value described in the text. The points are experimental and the
curve is calculated using the global fit results, eqs.
(\ref{ex13}), with the population balance parameter fixed at $\chi
=0$. We have not represented the imaginary part
$Im\left(\underline{\mathcal{V}}_B\right)(I)$ which is very small ($<
0.03$) and very sensitive to $\chi$. \label{fig10}}
\end{center}
\end{figure}

With a HMW coil current $I$ and a compensator coil current $I_C$, optimum compensation of the linear part of the Zeeman phase shift is obtained with $I_C \approx \left|I\right|/3$. When $\left|I\right|> 15$ A, it is impossible to use $I_C>5$ because the compensator coil temperature rises too much and, then, we have used $I_C=5$ A. Fig. \ref{fig10} presents the relative visibility $\mathcal{V}_B$ as a function of $I$. Comparison with fig. \ref{fig9} proves the efficiency of the compensator: when $I_C=0$, $\mathcal{V}_B (I)$ vanishes for $I\approx 6$ A while, with the compensator in operation, it remains larger than $80$\% if $\left|I\right|\leq 12$ A and vanishes only for $\left|I\right|\approx 18$ A. The revival observed for $I=23$ A, with a relative visibility near $-70$\% and a phase shift close to $\pi$, is explained in \cite{LepoutrePhD11}.

In order to have the best estimate of the Zeeman phase
shifts induced during the HMW effect measurements, we performed a
single global fit of all the data recorded while testing the effects
of magnetic fields. This data set was collected during the HMW
effect measurements using both coils (with $I_C$ related to $I$ for optimum compensation) as well as during calibration
measurements using either both coils (with different relative tuning of $I$ and $I_C$) or only one coil. As introduced in HMWI, the Zeeman phase shifts are calculated using $J_1 = A_{J1} \left|I-I_{0}
\right|+ A_{J1,C} \left|I_{C}-I_{0,C} \right| + J_{0,I+C}$, with $J_{0,I+C} = J_0 - A_{J1} \left|I_{0}\right| - A_{J1,C}
\left|I_{0,C}\right|$. The data set for $\mathcal{V}_B \left(I,I_C \right)$ and
$\varphi_B \left(I,I_C \right)$ includes about 150 data points which belong to $31$ series corresponding to slightly different laser frequencies and a different $\chi$ value is fitted for each series. Here are the fitted values of $J_0$, $I_0$, $A_{Ji}$, $I_{0,C}$ and $A_{J1,C}$ provided by this global fit:

\begin{eqnarray}
\label{ex13}
J_0     &=&   -0.61 \pm 0.01 \mbox{ rad} \nonumber \\
I_{0}   &=&    0.31 \pm 0.03 \mbox{ A} \nonumber \\
A_{J1}  &=&   -0.430 \pm 0.005 \mbox{ rad/A} \nonumber \\
A_{J2}  &=& \left(-662 \pm 5\right)\times 10^{-5} \mbox{ rad/A}^2 \nonumber \\
A_{J3}  &=& \left(-180\pm 5 \right)\times 10^{-6} \mbox{ rad/A}^3  \nonumber \\
I_{0,C} &=& \left( 22  \pm 9\right) \times 10^{-3} \mbox{ A} \nonumber \\
A_{J1,C}&=& 1.43 \pm 0.015 \mbox{  rad/A}
\end{eqnarray}

%%%%%%%%%%%%%%%%%%%%%%%%%%%%%%%%%%%%%%%%%%%%%%%%%%%%%
%%%%%%%%%%%%%%%%%%%%%%%%%%%%%%%%%%%%%%%%%%%%%%%%%%%%%%

\section{Data set for the HMW phase measurement and raw results}
\label{data}

%%%%%%%%%%%%%%%%%%%%%%%%%%%%%%%%%%%%%%%%%%%%%%%%%%%%%
\begin{figure}[t]
\begin{center}
\includegraphics[height= 6 cm]{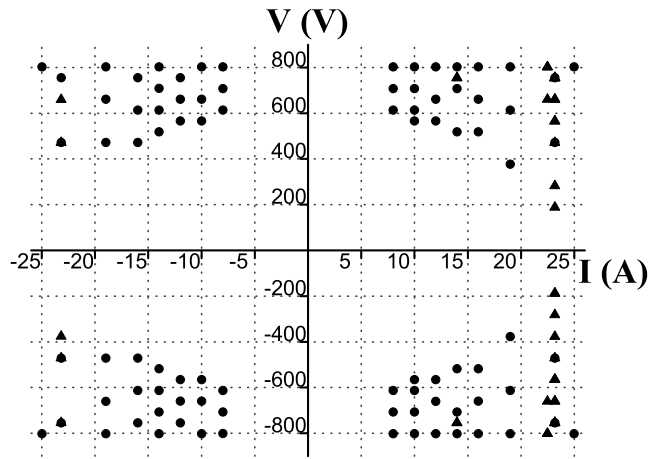} \caption{Data set
collected for the measurement of the HMW phase. Each run is
represented in the $I$,$V$ plane by a triangle (respectively a
bullet) for a 4-field (respectively 6-field) experiment.
\label{fig11}}
\end{center}
\end{figure}
%%%%%%%%%%%%%%%%%%%%%%%%%%%%%%%%%%%%%%%%%%%%%%%%%%%%%%

Fig. \ref{fig11} presents the data set collected for the HMW phase
measurement. As $\varphi_{HMW}$ is very small and proportional to the
$VI$ product, we have chosen to record data with large values either
of $V$ or of $I$, so that we have no data point near the origin.

\begin{figure}[t]
\begin{center}
\includegraphics[width= 8 cm]{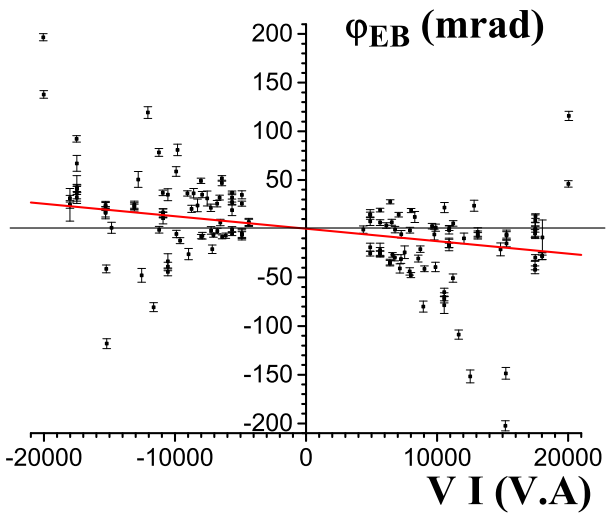}
\caption{(color online) Measured values of $\varphi_{EB}(V,I)$ given by eq. (\ref{a1}) as a function of the $VI$ product measured in VA. The (red) full line represents the expected value of the HMW phase, $\varphi_{HMW}(V,I)= - (1.28 \pm 0.03) \times 10^{-6} VI$ rad.  \label{fig12}}
\end{center}
\end{figure}

The measured values of the phase shift $\varphi_{EB}(V,I)$ are
plotted as a function of the $VI$ product in fig. \ref{fig12}: these
results do not agree with the predicted variations of $\varphi_{HMW}$
and we explain this discrepancy by stray phase shifts which appear when the
electric and magnetic fields are simultaneously applied. The origin
of these stray phases have been explained on general grounds in HMWI
and the detailed calculation is presented in the appendix of the
present paper. We are going to test these calculations first on the
relative visibility $\mathcal{V}_{EB}(V,I)$ and afterwards on the
phase shift $\varphi_{EB}(V,I)$. The various stray effects differ by their symmetry with respect to the reversal of the electric and/or magnetic fields and, in order to test separately these effects, it is necessary to extract the even/odd
parts of these quantities with respect to field reversals by combining
measurements for opposite $V$ or $I$ values. For any quantity
$f(V,I)$, the mean $\mathcal{M}_Xf(V,I)$ and the half-difference
$\Delta_X f(V,I)$ for opposite values of $V$ (then $X=E$) or of $I$ (then
$X=B$) are equal to:

\begin{eqnarray}
\label{a4}
\mathcal{M}_Ef(V,I)&=& \left[f(V,I) + f(-V,I)\right]/2 \nonumber \\
\Delta_Ef(V,I)&=& \left[f(V,I) -f(-V,I)\right]/2 \nonumber \\
\mathcal{M}_Bf(V,I)&=& \left[f(V,I) + f(V,-I)\right]/2 \nonumber \\
\Delta_Bf(V,I)&=& \left[f(V,I) - f(V,-I)\right]/2
\end{eqnarray}

\noindent Most experiments were done with 6 field configurations and they provide simultaneous measurements of $\mathcal{V}_{EB}(V,I)$ and $\varphi_{EB}(V,I)$ for opposite voltages, with exactly the same current $I$ and the same value of the population unbalance parameter $\chi$: we thus have very sensitive tests of the effects of electric field reversal.

%%%%%%%%%%%%%%%%%%%%%%%%%%%%%%%%%%%%%%%%%%%%%%%%%%%%%
%%%%%%%%%%%%%%%%%%%%%%%%%%%%%%%%%%%%%%%%%%%%%%%%%%%%%%

\section{Some experimental tests of the effects of stray phases}
\label{test}

We are going to test the predictions of the calculations described in the appendix of the present paper.

\subsection{Tests involving the fringe visibility}

Following eq. (\ref{a6}), four combinations of
$\mathcal{V}_{EV} (V,I)$ separate the contributions of the four
$D_{\pm,\pm}(V,I)$ terms. However, as shown by eq.
(\ref{a14}), the quantity $\mathcal{V}_{EB}$ also
includes a contribution due to the Aharonov-Casher effect in the
corresponding $(V,0)$ field configuration. It is given by the term
$D_{AC,B0}(V)/D_{0,B0}$, with a value close to $1.1$\%
for $V=800$ V. Because it involves the AC phase, this effect is an
odd function of the voltage $V$. We eliminate this contribution by using the measured values of $\mathcal{V}_E (\pm V)$
to calculate $\Delta_E\mathcal{V}_{E}(V)$, from which we deduce a corrected fringe visibility given by
$\mathcal{V}_{EB}^{'}(V,I) = \mathcal{V}_{EB}(V,I)/\left( 1 -
\Delta_E\mathcal{V}_{E}(V)\right)$. This quantity is now simply expressed by eq. (\ref{a3}) (obviously, this correction is necessary only when studying $V$-odd terms).

The variations of  $\mathcal{M}_B\Delta_E\mathcal{V}_{EB}^{'}(V,I)$
give a test of the $\sum D_{-,+}$ term. We do not plot these results
here because all the values are very small, in the $(-1\mbox{ to }
+5)\times 10^{-3}$ range, with error bars near $\pm 2\times 10^{-3}$,
with one exception, for $I=\pm 19$ A (the visibility is then very low
and some approximations of our calculations of the appendix are no more valid). The variations of
$\Delta_B\mathcal{M}_E\mathcal{V}_{EB}(V,I)$ give a test of the $\sum
D_{+,-}$ term. We do not plot these results here because all the
values are also very small, in the $(-2 \mbox{ to } +4)\times
10^{-3}$ range, with error bars near $\pm 2\times 10^{-3}$. These two
results prove that the $\sum D_{-,+}$ term and the $\sum D_{+,-}$
term are very small, in good agreement with our calculations which
predict that these terms should vanish if the contact potential terms
are negligible.

Having verified that the $D_{-,+}$ terms are negligible,
the quantity $\Delta_E \mathcal{V}_{EB}^{'}(V,I)$ reduces to $\sum
D_{-,-}/D_0$ (see eq. (\ref{a5})). The leading terms of $D_{-,-}$ given by eq. (\ref{Aa6}) are proportional to the Aharanov-Casher phase:

\begin{eqnarray}
\label{a15} \Delta_E \mathcal{V}_{EB}^{'}(V,I) \approx -\frac{\sum
\varphi_{AC} \sin \left( \phi_Z\right) }{\sum \cos \left( \phi_Z\right)
}
\end{eqnarray}

\begin{figure}[t]
\begin{center}
\includegraphics[width= 8 cm]{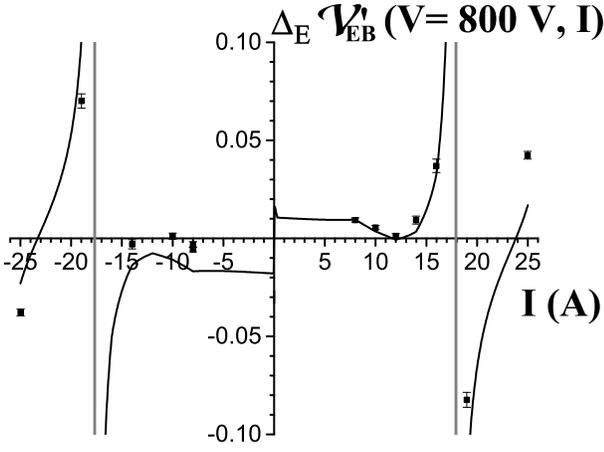} \caption{Plot of
$\Delta_E\mathcal{V}_{EB}^{'}(V= 800 \mbox{ V},I)$ as a function of the current
$I$: the measured data points (squares) are compared to the result of our model (full line). The visibility, proportional to $D_0 = \sum \cos \left(\phi_Z\right)$ vanishes when $I\approx\pm 18$ A indicated by vertical
lines: this induces a divergence of the prediction of our model, which uses a first-order calculation in $D_{\pm,\pm}/D_0$.
Our model explains well the main variations of $\Delta_E \mathcal{V}_{EB}^{'}(V,I)$, even if some imperfections appear clearly.
\label{fig13}}
\end{center}
\end{figure}

\begin{figure}[t]
\begin{center}
\includegraphics[width= 8 cm]{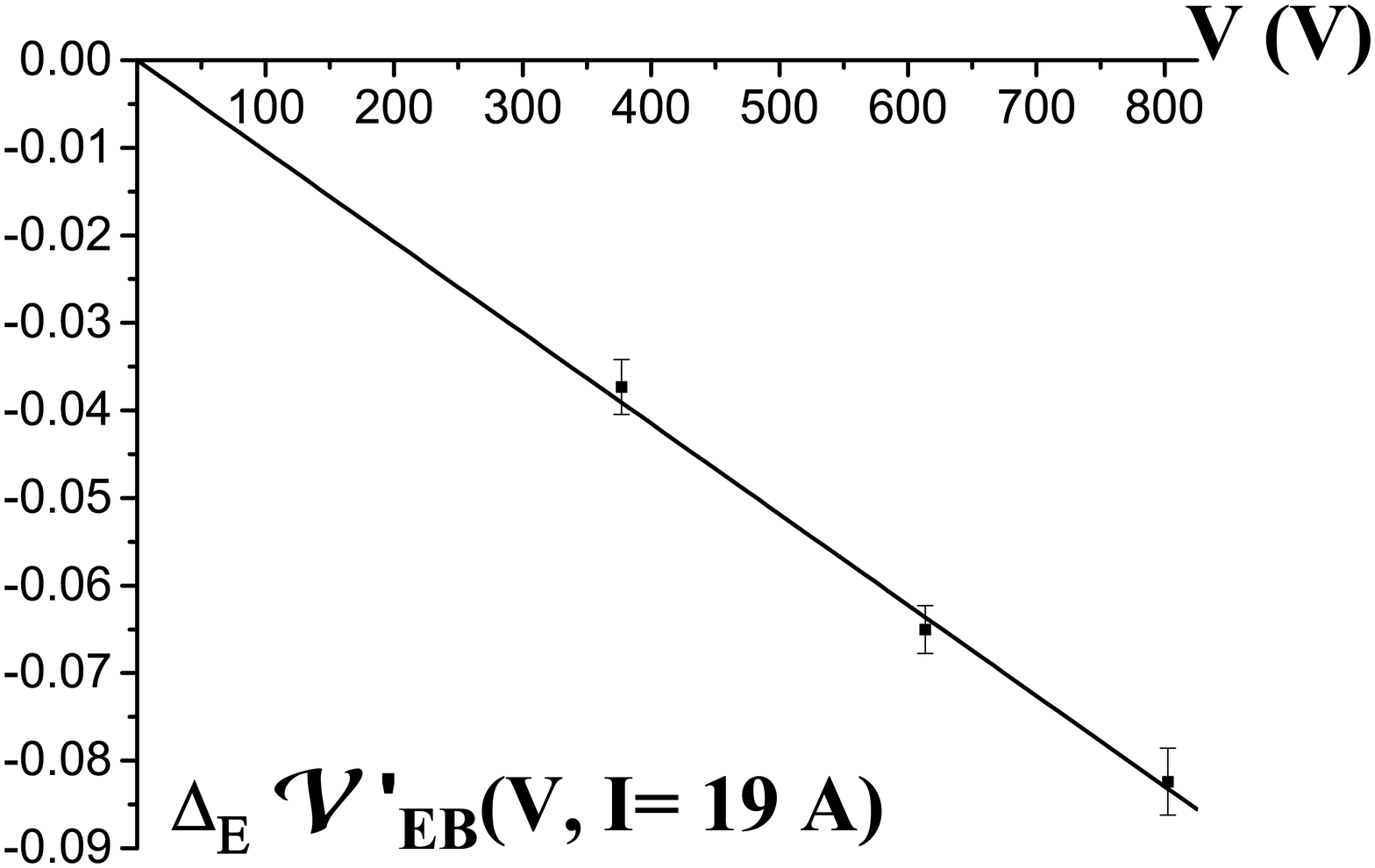} \caption{Plot of
difference of visibility for opposite $V$ values,
$\Delta_E\mathcal{V}_{EB}^{'}(V,I=19\mbox{ A})$, as a function of $V$. The
current value $I=19$ A, chosen  close to the cancelation of $D_0 =
\sum \cos \left( \phi_Z\right)$, enhances the sensitivity of the
visibility to the AC phase. The measured values (squares) are
well represented by a linear function of $V$, as predicted by our
model. \label{fig14}}
\end{center}
\end{figure}

\noindent  Thanks to our knowledge of the Zeeman phases (eqs. (\ref{ex13})), we can evaluate all the terms of
eq. (\ref{a15}) and we compare its prediction to our measurements in
fig. \ref{fig13} and fig. \ref{fig14}. The good agreement, obtained without any fitted parameter, proves that the
dominant $V$-odd effect is due to the AC phase shift, and
confirms the validity of our calculations.

%%%%%%%%%%%%%%%%%%%%%%%%%%%%%%%%%%%%%%%%%%%%%%%%%%%%%
%%%%%%%%%%%%%%%%%%%%%%%%%%%%%%%%%%%%%%%%%%%%%%%%%%%%%%

%%%%%%%%%%%%%%%%%%%%%%%%%%%%%%%%%%%%%%%%%%%%%%%%%%%%%
%%%%%%%%%%%%%%%%%%%%%%%%%%%%%%%%%%%%%%%%%%%%%%%%%%%%%%

\subsection{Tests involving the fringe phase}

We first discuss the combination
$\mathcal{M}_B\Delta_E\varphi_{EB}(V,I)$ given by:

\begin{eqnarray}
\label{a16} \mathcal{M}_B\Delta_E\varphi_{EB}(V,I)&=& -\frac{\sum
N_{-,+} }{D_0}
\end{eqnarray}

\begin{figure}[t]
\begin{center}
\includegraphics[width= 8 cm]{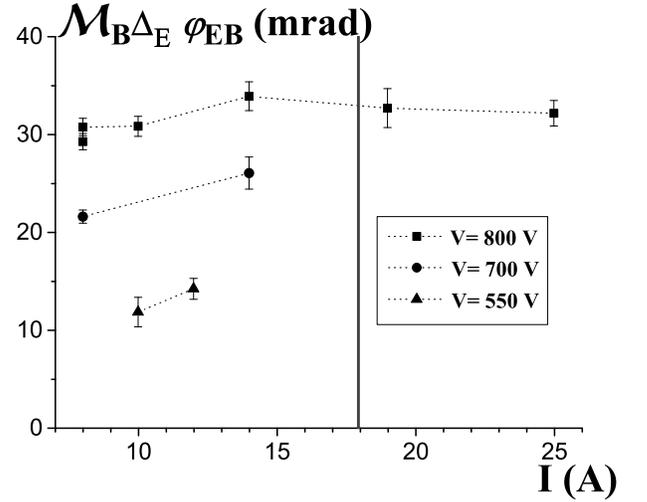} \caption{Plot of the
quantity $\mathcal{M}_B\Delta_E\varphi_{EB}(V,I)$ in radians as a
function of the current $I$ for several values of the voltage $V$
applied to the capacitors. The dotted lines are simply connecting
values measured for the same $V$ values. The vertical line for
$I\approx 18$A indicates the place where the fringe visibility
vanishes. \label{fig15}}
\end{center}
\end{figure}

\begin{figure}[t]
\begin{center}
\includegraphics[width= 8 cm]{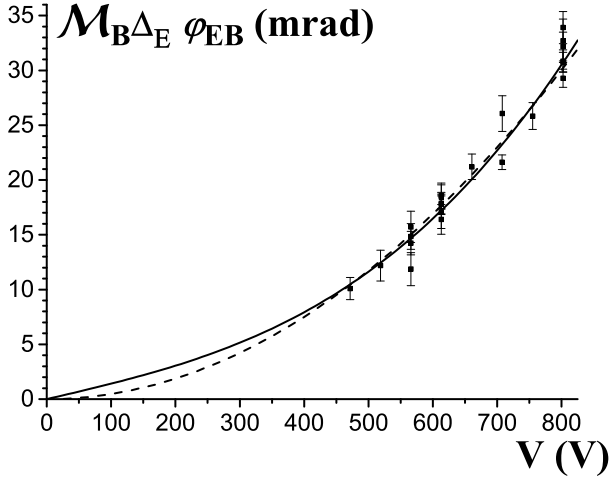} \caption{Plot of the
quantity $\mathcal{M}_B\Delta_E\varphi_{EB}(V,I)$ as a
function of the voltage $V$ applied to the capacitors for all the
values of the current $I$ in the HMW coil. The dotted line is a fit
with a single $V^3$ term while the full line is a fit with a term in $V$ and a term in $V^3$. \label{fig16}}
\end{center}
\end{figure}

\noindent As $N_{-,+}$ is non-zero only if contact potentials are not
negligible, we expected this quantity to be negligible.
$\mathcal{M}_B\Delta_E\varphi_{EB}$ is plotted as a
function of $I$ on fig. \ref{fig15} and as a function of $V$ on fig.
\ref{fig16}. These experimental results are surprising:
$\mathcal{M}_B\Delta_E\varphi_{EB}(V,I)$  is almost independent of the
current $I$ and it rapidly increases with the voltage $V$. The
measured values are well fitted as the sum of two terms, one term in $V$ and
one in $V^3$ (odd powers of $V$ have been chosen because this quantity is $V$-odd).

Contact potentials can in principle explain non zero values of $\mathcal{M}_B\Delta_E\varphi_{EB}$ (the calculation is made in ref. \cite{LepoutrePhD11}), but the predicted effect depends of the current $I$ with divergences similar to those visible in fig. \ref{fig13}, in complete disagreement with the measurements plotted in fig. \ref{fig15}. Moreover, the observed
magnitude of $\mathcal{M}_B\Delta_E\varphi_{EB}(V,I)$ would require
values of contact potentials that are ruled out by the measurements of $\mathcal{V}_E (V)$ and $\mathcal{V}_{EB} (V,I)$
previously presented. This effect is strange because
$\varphi_{EB}(V,I)$ given by eq. (\ref{a1}) is already a
difference of phase shifts measured with and without the magnetic
field, so that $\mathcal{M}_B\Delta_E\varphi_{EB}(V,I)$ must vanish when
the applied magnetic field goes to zero: as a consequence, the
independence of $\mathcal{M}_B\Delta_E\varphi_{EB}(V,I)$ with the
current $I$ cannot extend to $I\rightarrow 0$. However, if the
transition occurs for instance when the laboratory field and the HMW field are
comparable in magnitudes, it should be observed with a
current $I$ of the order of $0.1$ A, a range of $I$-values we have
not studied.

We have investigated several possible explanations which revealed
unsatisfactory for different reasons: usually, either the symmetry
with respect to $V$ and $I$ reversals or the order of magnitude of
the observed phase are not in agreement with our observations.
Moreover, most explanations cannot explain why the
effect is sensitive to the presence of the magnetic field but
independent of its value in the studied range in figure \ref{fig15}.
We will not discuss here these failed explanations, for lack of
space. The origin of this phase shift remains mysterious but thanks
to its independence with regards to $I$, it can be
easily eliminated by combining data with opposite $I$-values.

%%%%%%%%%%%%%%%%%%%%%%%%%%%%%%%%%%%%%%%%%%%%%%%%%%%%%
%%%%%%%%%%%%%%%%%%%%%%%%%%%%%%%%%%%%%%%%%%%%%%%%%%%%%%

%%%%%%%%%%%%%%%%%%%%%%%%%%%%%%%%%%%%%%%%%%%%%%%%%%%%%
%%%%%%%%%%%%%%%%%%%%%%%%%%%%%%%%%%%%%%%%%%%%%%%%%%%%%%

We now discuss the quantity
$\Delta_B\mathcal{M}_E\varphi_{EB}(V,I)$ which vanishes if contact potentials are negligible. If they are taken
into account, this quantity is given by \cite{LepoutrePhD11}:

\begin{eqnarray}
\label{a17} \Delta_B\mathcal{M}_E\varphi_{EB}(V,I) \approx \frac{\sum
\varphi_{AC} \left< \delta \varphi_{S,c}\delta\varphi_Z\right>\cos
\left( \phi_Z\right) }{\sum \cos \left( \phi_Z\right) }
\end{eqnarray}

\begin{figure}[t]
\begin{center}
\includegraphics[width= 8 cm]{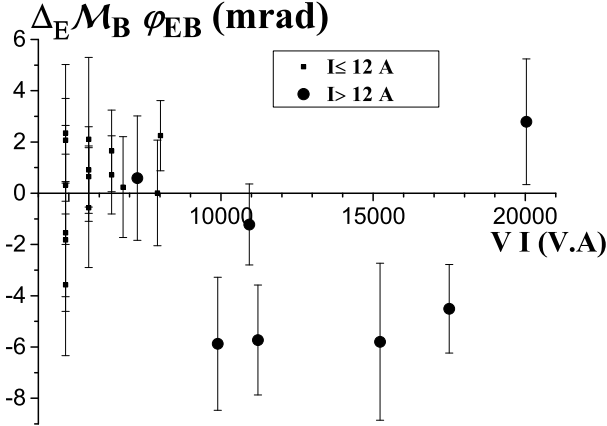} \caption{Plot of the
quantity $\Delta_B\mathcal{M}_E\varphi_{EB}(V,I)$  as a
function of the $VI$ product: different symbols depending if
$\left|I\right|\leq 12 $ A or $\left|I\right|> 12 $ A. \label{fig17}}
\end{center}
\end{figure}

\noindent Because of the presence of a contact potential term $\delta \varphi_{S,c}$,  $ \Delta_B\mathcal{M}_E\varphi_{EB}(V,I)$ is expected to be small, and we have not included higher order terms in eq. (\ref{a17}), because they should be even smaller. The measured values of $\Delta_B\mathcal{M}_E\varphi_{EB}(V,I)$ are plotted in fig.
\ref{fig17}, with different symbols for data points depending if
$\left|I\right|$ is smaller or larger than $12$ A. When
$\left|I\right|\leq 12 $ A, the measured values are very small and
compatible with $0$: in this range of $I$ values, the Zeeman phases
$\phi_Z$ are small thanks to the compensator, the systematic effects
are weak and the approximations done in our model are good. When
$\left|I\right|> 12 $ A, the Zeeman phases increase
rapidly with $\left|I \right|$, and several effects decrease
the accuracy of our model. First, the polynomial expansion of the Zeeman
phases in powers of  the current $I$  is poorly convergent for some sublevels (see HMWI) while the systematic effects are very sensitive to the value of the Zeeman phases.
Secondly, with increasing Zeeman phases, the systematic effects which
involve the dispersion $\delta \varphi_Z$ increase (this point is discussed
below). Finally, increasing Zeeman phases induce a rapid decrease of the
visibility which cancels for $I\approx18$ A and higher
order terms in $N_i/D_0$ or $D_i/D_0$ are no more negligible.

%%%%%%%%%%%%%%%%%%%%%%%%%%%%%%%%%%%%%%%%%%%%%%%%%%%%%
%%%%%%%%%%%%%%%%%%%%%%%%%%%%%%%%%%%%%%%%%%%%%%%%%%%%%%

\begin{figure}[t]
\begin{center}
\includegraphics[width= 8 cm]{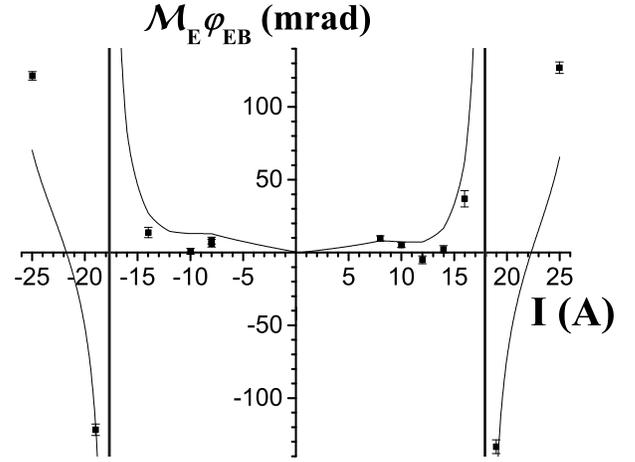} \caption{\label{fig17SL18} Plot of the
quantity $\mathcal{M}_E\varphi_{EB}(V,I)$ for $V=800$ V, as a function of the
current $I$. The points are measured values, the line is a best fit
for all the measurements of $\mathcal{M}_E\varphi_{EB}$ (see
discussion). As for fig. \ref{fig13}, vertical gray lines indicate
cancelation of the visibility.}
\end{center}
\end{figure}

The tests on the fringe phase presented up to now have detected stray phase shifts not larger than $35$ mrad. We end this part by
considering the quantity $\mathcal{M}_E\varphi_{EB}$, which includes the largest stray phase shifts. $\mathcal{M}_E\varphi_{EB}(V,I)$ is an even function of the current $I$, because we have  just shown that
$\Delta_B \mathcal{M}_E \varphi_{EB} (V,I) = \left[\mathcal{M}_E\varphi_{EB} (V,I) - \mathcal{M}_E\varphi_{EB}
(V,-I)\right]/2$ is negligibly small. $\mathcal{M}_E\varphi_{EB}$ is given by:

\begin{eqnarray}
\label{AjoutSL} \mathcal{M}_E \varphi_{EB} (V,I) &=& -\frac{\sum N_{+,+}}{D_0} \nonumber \\
&\approx& - \frac{\sum \left< \delta \varphi_{S,g} \delta \varphi_Z \right> \sin (\phi_Z)}{\sum \cos (\phi_Z)}
\end{eqnarray}

\noindent where we have neglected higher-order terms (see eq. (\ref{Aa6})). The measured values of
$\mathcal{M}_E\varphi_{EB}(V,I)$ for $V=800$ V are plotted in fig.
\ref{fig17SL18}. The Stark phase dispersion $\delta \varphi_{S,g}
(y)\propto V^2$ has been characterized thanks to the study of
$\mathcal{V}_E(V)$ (part \ref{elec3}). The evaluation of the
variations of $\delta \varphi_Z (y)$ with $I$ is done at the expense
of a supplementary approximation, assuming a rectangular profile for
the field of the HMW coil along the atom trajectory (compare fig.
\ref{fig2}). It then becomes possible to perform a fit of the measured values of $\mathcal{M}_E\varphi_{EB}$
(taking into account the terms neglected in eq.
(\ref{AjoutSL})). The result of this fit is also shown in figure
\ref{fig17SL18}: a good agreement is found for the behavior of this
quantity, and the fitted parameter values are compatible with the
expected dispersion of $\delta \varphi_Z(y)$ along the atomic beam
height according to the calculations of the magnetic field. This result confirms the importance of the spatial dispersion $\delta \varphi_Z (y)$ of the Zeeman phase shifts and it proves that the main systematic effects are due to these spatial phase dispersions.

%%%%%%%%%%%%%%%%%%%%%%%%%%%%%%%%%%%%%%%%%%%%%%%%%%%%%
%%%%%%%%%%%%%%%%%%%%%%%%%%%%%%%%%%%%%%%%%%%%%%%%%%%%%%

\subsection{ Conclusion concerning systematic effects}

Here are the main results of our study of these effects:

\begin{itemize}

\item the effects of the spatial dispersions $\delta\varphi_{S,g}$
and $\delta \varphi_Z$ of the Stark and Zeeman phase shifts
respectively are well identified;

\item the effects of the the dispersion $\delta\varphi_{S,c}$ of the
Stark phase due to contact potentials appear to be below our
experimental sensitivity;

\item our model provides a qualitative understanding of the systematic
effects for all values of the current $I$. The visibility decreases
rapidly and vanishes for $\left|I\right|\approx 18 $ A: this
circumstance has been used to enhance the sensitivity to certain
terms but clearly, as soon as in $\left|I\right|> 12 $ A, our
model describing the systematic effects is less accurate.

\item as the visibility presents a revival for $\left|I\right|\approx 23 $ A with the Zeeman phases $\phi_Z$ being close to $\pm \pi$, we have made several series of measurements in this range of $I$-values but we cannot expect our model to be accurate;

\item we have observed an unexpected phase shift which is independent
of the current $I$ in the range $8-23$ A and which is odd with
respect to $V$-reversal. We presently have no explanation
for this effect and we continue our investigations on
its possible sources. It may be either a systematic effect forgotten in our analysis or a fundamental physical effect,
for instance such as the effects discussed by J. Anandan
\cite{AnandanPRL00} but, as far as we can judge, these fundamental
effects are either too small or they have not the correct symmetry
with respect to $V$ and $I$.

\end{itemize}

\section{Measurement of the HMW phase}
\label{HMWm}

We now use our knowledge of the stray phase shifts in order to
eliminate their contributions to the measurement of the HMW phase.
The HMW phase $\varphi_{HMW}$ is proportional to the $VI$
product, i.e. it is odd with respect to $V$- and
$I$-reversals. The main contribution in the stray phase
shifts on the measurements of $\varphi_{EB}$ are even with respect to
$V$ and $I$, but because of the existence of a $V$-odd phase of
unknown origin, we choose to use the $I$-odd character of
$\varphi_{HMW}$ to cancel the maximum amount of systematic effects.
Accordingly, we plot the quantity $\Delta_B\varphi_{EB}(V,I)$ as a
function of the $VI$ product. We have used different symbols for the
measurements depending if $\left|I\right|$ is smaller or larger than
$12$ A and we have made separate fits of these two sets of data using
$\Delta_B\varphi_{EB}(V,I) = \alpha VI + \beta$.

\begin{eqnarray}
\label{sm4}
\alpha &=& \left( -1.94 \pm 0.06 \right)\times 10^{-6}  \mbox{ rad/VA } \nonumber \\
\beta  &=& \left( 7 \pm 4 \right) \times 10^{-4} \mbox{ rad }       \nonumber \\
\mbox{  if} && \left|I\right|\leq 12 \mbox{ A}
\end{eqnarray}

\noindent and

\begin{eqnarray}
\label{sm5}
\alpha &=& \left( -2.16 \pm 0.14 \right)\times 10^{-6} \mbox{ rad/VA } \nonumber \\
\beta  &=& \left( -26 \pm 19 \right) \times 10^{-4} \mbox{ rad }     \nonumber \\
\mbox{  if} && \left|I\right|> 12 \mbox{ A}
\end{eqnarray}

\noindent In both fits, the intercept $\beta$ for $VI=0$ is
compatible with a vanishing value. The error bar on the slopes
$\alpha$ is substantially smaller when $\left|I\right|\leq 12$ A than
when $\left|I\right|> 12$ A: this is visible on the data which is
more dispersed when  $\left|I\right|> 12$ A. For both
fits, the fitted slopes are larger (in modulus) than the predicted
value $\varphi_{HMW}  \left(V, I \right)/(VI) = - (1.28 \pm 0.03)
\times 10^{-6} $ rad/VA. The discrepancy is equal $52$\% if
$\left|I\right|\leq 12$ A and $69$\% if $\left|I\right|> 12$ A. Our
model predicts that there are two contributions to
$\Delta_B\varphi_{EB}(V,I)$:

\begin{eqnarray}
\label{a18}
\Delta_B\varphi_{EB}(V,I) &=& \varphi_{HMW} - \frac{N_{-,-}}{D_0}
\end{eqnarray}

\noindent $N_{-,-}$ given by equation \ref{Aa6} is the
product of the AC phase by correlation terms. Thanks to the knowledge
of the experimental defects, it is possible to evaluate all the terms
involved in $N_{-,-}/D_0$. The only quantities which are not directly
measured are the correlations $ \left<\delta \varphi_d \delta
\varphi_Z\right>$ and $ \left<\delta \varphi_d \delta
\left(\varphi_Z\right)^2\right>$ which are evaluated from the
measurement the correlation with $\delta \varphi_S$ replacing $\delta
\varphi_d$, assuming that both effects are linear functions of $y$.
The calculated value of $N_{-,-}/D_0$ never exceeds $3$
mrad for the data set with $\left|I\right|\leq 12$ A,
and we have made this correction to get $\varphi_{final}(V,I)$ which is
plotted  in fig. 3 of our letter \cite{LepoutrePRL12}. The fitted
slope $\varphi_{final}(V,I)/VI = (-1.68 \pm 0.07)\times 10^{-6}$  rad/V.A is
still too large but the discrepancy with the theoretical value is reduced to $31$\%.

%%%%%%%%%%%%%%%%%%%%%%%%%%%%%%%%%%%%%%%%%%%%%%%%%
\begin{figure}[h]
\begin{center}
\includegraphics[width= 8 cm]{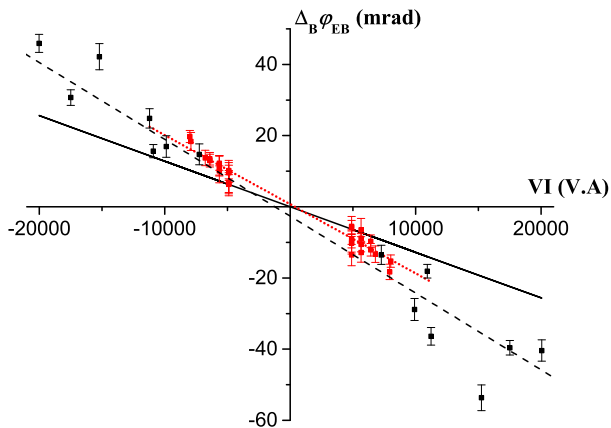} \caption{(color online) Measured values
of $\Delta_B\varphi_{EB}(V,I)$ as a function of the $VI$ product. The
data points with $\left|I \right| \leq 12$ A are plotted in red as
well as their fit represented by a dotted line. The data points with
$\left|I \right| > 12$ A are plotted in black as well as their fit
represented by a dashed line. The expected dependence of
$\varphi_{HMW}(V,I)$ with the $VI$ product is represented by a black full line.
\label{FigDBPhiAEB}}
\end{center}
\end{figure}
%%%%%%%%%%%%%%%%%%%%%%%%%%%%%%%%%%%%%%%%%%%%%%%%%%%%%%

\section{Conclusion}
\label{Conc}

\subsection{Some remarks on the present experiment}

We have described a measurement of the He-McKellar-Wilkens
topological phase by atom interferometry. This experiment was
feasible with our atom interferometer because the interferometer arms
are well separated in space and the interferometer signal is intense,
with a large fringe visibility, near $70$\%. The arm separation is
needed in order to insert a septum between the two interferometer
arms without any degradation of the signal. The signal intensity and
the large value of the fringe visibility both contribute to enhance
the phase sensitivity: its practically achieved value near $25$
mrad/$\sqrt{\mbox{Hz}}$ is needed for the present measurement. The
HMW phase shift is rather small, at most $27$
mrad under our experimental conditions, and appears as
the combination of four phase measurements for which
$2000$ s of data recording were needed to reduce the uncertainty
near $3$ mrad.

The analysis of the experiment revealed more complex than expected,
because of stray phases. The complexity of the signal is due to several
factors:

\begin{itemize}

\item the signal is the sum of the contributions of 8 sublevels which
are not exactly in phase because of the Zeeman phase shifts due to the
slightly different values of the magnetic field on the two interferometer arms;

\item we have built a compensator coil to produce an opposite gradient
of the magnetic field at another place in the interferometer. The use
of this compensator has been very fruitful as it has enabled us to
apply substantially higher fields with a limited loss of fringe
visibility. However the compensator produces a low field, so that it
can correct only the part of the phase shifts due to linear Zeeman
effect;

\item the weights of the various $F, m_F$ sublevels are functions of
the laser frequency and power density in the standing waves used for
atom diffraction. We had to control these parameters rather tightly
in order to keep these weights almost equal and constant;

\item the main phase shifts (diffraction phase shift, Stark and Zeeman
phase shifts) present a dispersion with the atomic trajectory
described in our calculations by the $y$-coordinate. In the presence
of several dispersed phase shifts, the visibility of the contribution
of a given sublevel to the total fringe signal is better or worse,
depending if the dispersions of the different phase shifts subtract or add their
effects.

\end{itemize}

We have developed a model taking into account all these effects and
this model has been very successful in explaining the variations of
the observed phase shifts and visibility with the capacitor voltage
$V$ and the HMW coil current $I$. However, an extra-phase has been
observed and characterized: this phase is odd with the
capacitor voltage $V$; it behaves roughly like $V^3$; it appears only when the
magnetic field is applied but its value is independent of the
magnetic field magnitude in a wide range. We continue our
investigations to understand the effect which produces this phase.
By combining measured phase shifts with opposite values of the current
$I$, we have eliminated this phase and we have obtained a first
measurement of the HMW phase shift. The observed effect is larger
than its expected value  by $69$\% if we use all the collected data
points and only by $52$\% if we consider only the data points with
$\left|I \right|\leq12$ A. Finally, there is a small stray contribution of
the Aharonov-Casher phase to the measured phase shift, and
using our model, it was possible to evaluate this contribution and to
correct the measured values accordingly. The discrepancy
between our corrected measurements and the expected HMW phase shift
is then reduced to $31$\%.

\subsection{Possible improvements of this experiment}

It is necessary to improve this experiment in order to reduce the
uncertainty on the HMW phase-shift. Here are the main possibilities:

\begin{itemize}

\item reduction of stray effects by a better construction of the HMW
interaction region. The present construction has two main defects:
the difference of the capacitor thicknesses varies with the
$y$-coordinate and the septum does not coincide with the
symmetry plane of the HMW coils. The construction of capacitors with
a better controlled geometry is probably possible but quite
difficult, because of the need of using a stretched septum. A better
centering of the septum with respect to the HMW coils is probably
rather easy and this would reduce substantially the Zeeman phase
shifts which are the largest source of complication.

\item  reduction of stray effects by optical pumping of the atomic
beam. If all the atoms are in one $F,m_F$ sublevel only, the signal
is no more an average on the hyperfine sublevels
populations. Moreover, the trajectory-averaged Zeeman phase shift
can be exactly canceled by the compensator if the pumping is done in
the $F=2$, $m_F=+2$ (or $-2$) sublevel for which Zeeman effect is
exactly linear. As a consequence, this arrangement, which should reduce most
of the stray phase shifts, is feasible with minor modifications
of our setup and experiments are in progress.

\item reduction of stray effects by using another atom: this requires
the development of a completely new atom interferometer with
separated arms. Most of the difficulties are due to the paramagnetic
character of lithium and an atom with a $^1$S$_0$ non-degenerate
ground state (i.e. with a zero nuclear spin) would be ideal
because there would be no Zeeman phase shift and no Aharonov-Casher
phase shift. We may consider either a thermal beam of a light atom or
a laser-cooled atomic source. In the case of a thermal beam, the most
obvious choice is ground state helium, with which a very nice
interferometer using diffraction by material gratings was developed
by J.P. Toennies and co-workers \cite{CroninRMP09}. Because helium
electric polarizability is small ($\alpha_{He} \approx
\alpha_{Li}/120$), larger electric and/or magnetic fields will be
needed. Among atoms which have been laser-cooled, magnesium, calcium,
strontium or ytterbium all have a $^1$S$_0$ ground state and at least one
stable isotope with a nuclear spin equal to $0$.
\end{itemize}

%%%%%%%%%%%%%%%%%%%%%%%%%%%%%%%%%%%%%%%%%%%%%%%%%%%%%%%%%%%%%%%%%%%%%%

\acknowledgments

We thank the laboratory technical staff for their help, A. Cronin for fruitful discussions, G. Tr\'enec, A. Miffre and M. Jacquey for all the work done on our atom interferometer. We are greatly indebted toward CNRS INP, ANR (grants ANR-05-BLAN-0094 and ANR-11-BS04-016-01 HIPATI) and R\'egion Midi-Pyr\'en\'ees for supporting our research.

%%%%%%%%%%%%%%%%%%%%%%%%%%%%%%%%%%%%%%%%%%%%%%%%%%%%%%%%%%%%%%%%

\section{Appendix: calculation of the fringe signal}

\label{App}

We describe here the main points of our calculation of the fringe
signal from which we deduce the stray phase and the fringe
visibility.

\subsection{Some simplifying assumptions}
\label{theory1}

The fringe phase $\varphi$  is the sum of the diffraction phase
$\varphi_d$, the Sagnac phase $\varphi_{Sagnac}$ due to Earth
rotation, the Stark phase $\varphi_S$, the Zeeman phase
$\varphi_{Z}(F,m_F)$, the HMW phase $\varphi_{HMW}$ and the
Aharonov-Casher phase $\varphi_{AC}(F,m_F)$:

\begin{eqnarray}
\label{t0} \varphi&=&  \varphi_d +  \varphi_{Sagnac} + \varphi_S +
\varphi_{Z}(F,m_F) \nonumber \\ && + \varphi_{HMW} + \varphi_{AC}(F,m_F)
\end{eqnarray}

\noindent $\varphi_d$, $\varphi_{HMW}$ and $ \varphi_{AC}(F,m_F)$ are
independent of the atom velocity $v$; $\varphi_{Sagnac}$ and
$\varphi_S$ vary like $1/v$ and $\varphi_{Z}(F,m_F)$ like $1/v^2$.
These velocity-dependent phases are small ($\varphi_{Sagnac}\approx
0.65$ rad,  $\left| \varphi_S \right| \lesssim 0.2$ rad and
$\left|\varphi_{Z}(F,m_F)\right|<2$ rad when $\left|I\right|< 18$ A - only these data points will be retained for the final
analysis). As the parallel speed ratio $S_{\|}$ of the lithium beam
is large, $S_{\|}\approx 9$, we may forget the velocity
average and, as a consequence, the Sagnac phase $\varphi_{Sagnac}$
which is constant. We consider the spatial dispersion of $\varphi_d$,
$\varphi_S$ and $\varphi_{Z}(F,m_F)$ only and we neglect this
dispersion for $\varphi_{HMW}$ and $\varphi_{AC}(F,m_F)$, because they
are small, $\left| \varphi_{HMW} \right|< 27$ mrad and
$\left|\varphi_{AC}(F,m_F)\right|< 70$ mrad, for our largest fields.
The total phase dispersion $\delta \varphi$ is the sum of three terms
only:

\begin{eqnarray}
\label{t01}
\delta \varphi &=&  \delta \varphi_d + \delta \varphi_S + \delta \varphi_Z
\end{eqnarray}

\noindent From now on, the $F,m_F$ dependence of $\varphi_Z$ and
$\varphi_{AC}$ is not explicit and, for $\varphi_d$, $\varphi_S$ and $\varphi_Z$, we note $\phi_X$ the spatial
average of $\varphi_X$ given by $\phi_X=\left< \varphi_X \right> =
\int dy P(y)  \varphi_X(y) $. The average over the $F,m_F$ sublevels
is taken with equal weights, $P(F,m_F) = 1/8$. This is a
good approximation because in the experiments devoted to the HMW
phase measurement, we have kept $\chi$ small
($\left|\chi\right| <0.03$) and randomly distributed
around $0$ (its main effect is to induce a supplementary dispersion
of our phase measurements). With these approximations, following the
discussion of section IV of HMWI, the signal due to one $F,m_F$
sub-level is given by:

\begin{eqnarray}
\label{Aa1}
I(F,m_F) &=& I_0 \left[1 + \mathcal{V}_m\left<\cos\left(\varphi_m\right)\right>\right]/8 \nonumber \\
\mathcal{V}_m &=& \mathcal{V}_0 \left[ 1 -\left< {\left(\delta \varphi\right)}^2/2\right> \right] \nonumber \\
\varphi_m &=& \phi- \left< {\left(\delta \varphi \right)}^3/6\right>\nonumber \\
\mbox{with  } \phi &=& \phi_d +\phi_S+\phi_{Z}(F,m_F)+ \varphi_{HMW}+ \varphi_{AC}(F,m_F) \nonumber \\
\mbox{and  } \delta \varphi &=& \delta\varphi_d +\delta \varphi_S+ \delta\varphi_{Z}(F,m_F)
\end{eqnarray}

\noindent If we neglect nuclear magnetism, the $F,m_F$ sublevels form
4 pairs with exactly opposite Zeeman energy shifts: three pairs of
levels with the same $m_F$ value and the pair $F=2, m_F = \pm 2$. We
label these pairs by an $m_F$ value going from $-1$ to $+2$ and we
note $\varphi_{Z}$ the value of $\varphi_{Z}(F,m_F)$ for the sublevel
$F=2,m_F$.

\subsection{Tutorial calculation}

\noindent Because of numerous terms, these calculations are rather
complicated and we first present a tutorial calculation in which we
cancel $\delta\varphi_d$ and $\varphi_{AC}(F,m_F)$, and
we forget the cubic term in $\delta \varphi$.  We first calculate the
signal $I_{pair}$ of a pair of sublevels:

\begin{eqnarray}
\label{Aa2} \left[\frac{I_{pair}}{I_{0}/4} -1\right]/ {\mathcal{V}}_0 & \approx & \left[1 -\frac{\left<\left(\delta \varphi_S\right)^2\right>+\left<\left(\delta \varphi_{Z}\right)^2\right>}{2}\right] \nonumber \\ &\times&
\cos\left(\phi_{Z}\right) \cos\left( \phi_d + \phi_S + \varphi_{HMW} -\theta \right) \nonumber \\
\mbox{with   } \tan \theta &\approx & \theta \approx \left<\delta \varphi_S\delta \varphi_{Z}\right> \tan\left(\phi_{Z}\right)
\end{eqnarray}

\noindent The important point is the phase shift $\theta$
proportional to the correlation term $\left<\delta \varphi_S\delta
\varphi_{Z}\right>$ and this effect is due to the fact that the
contributions of the two levels of the pair have different
visibility: the term $\left[ 1 -\left<\left(\delta \varphi_{S}+
\delta \varphi_{Z}\right)^2\right>/2\right]$ modifies these
visibility in a different way because the dispersions $\delta
\varphi_S$ and $\delta\varphi_{Z}$ have the same sign for one level
of the pair and opposite signs for the other one. Because of the
$\tan\left(\phi_{Z}\right)$ factor, $\theta$ is very sensitive to
$\phi_{Z}$ value, especially when $\phi_{Z}$ is close to $\pi/2$.

\subsection{Complete calculation}

If we remove the approximations done in the tutorial example, we get
the signal $I_{tot}$ which has a form analogous to equation
(\ref{Aa2}):

\begin{eqnarray}
\label{Aa3}\left[\frac{I_{tot}}{I_{0}} -1\right]/ {\mathcal{V}}_0 &\approx&  \frac{1}{4}\left[D \cos\left( \phi_d + \phi_S + \varphi_{HMW} \right) \right. \nonumber \\    &&+ \left. N \sin\left( \phi_d + \phi_S + \varphi_{HMW} \right)\right] \nonumber \\
&\approx &  \frac{\sqrt{D^2 + N^2}}{4} \cos\left( \phi_d + \phi_S + \varphi_{HMW} -\theta \right) \nonumber \\
\mbox{with   } \tan \theta &\approx & \theta \approx \frac{N}{D}
\end{eqnarray}

\noindent  The numerator  $N$ and the denominator $D$ of the fraction
giving $\theta $ are given by:

\begin{eqnarray}
\label{Aa4}D&=&  \left[1 -\frac{\left<\left(\delta
\varphi_d +  \delta \varphi_S\right)^2\right>}{2} \right] D_0 + D_Z
+ D_{+/-}\nonumber \\    N &=& \frac{\left<\left(\delta \varphi_d +
\delta \varphi_S\right)^3\right>}{6} D_0 + N_Z + N_{+/-}
\end{eqnarray}

\noindent with the following definitions:

\begin{eqnarray}
\label{Aa5}D_0&=& \sum \cos\left(\phi_{Z}\right) \nonumber \\  D_Z &=& \sum \left[-\frac{\left<\left(\delta \varphi_Z\right)^2\right>}{2} \cos\left(\phi_{Z}\right) \right. \nonumber \\ && + \left. \frac{\left<\left(\delta \varphi_Z\right)^3+ 3 \left(\delta \varphi_d\right)^2\delta \varphi_Z\right>}{6} \sin\left(\phi_{Z}\right)\right]\nonumber \\
D_{+/-} & =& \sum\left[ D_{+,+} + D_{-,+} + D_{+,-} + D_{-,-}\right]\nonumber \\
N_Z &=& \sum \left[\left<\delta \varphi_d \delta \varphi_Z\right> \sin\left(\phi_{Z}\right)+ \left<\delta \varphi_d\left( \delta \varphi_Z\right)^2\right> \cos\left(\phi_{Z}\right)\right] \nonumber \\
N_{+/-} & =& \sum \left[ N_{+,+} + N_{-,+} + N_{+,-} + N_{-,-}\right]
\end{eqnarray}

\noindent In these equations, $\sum$ is the sum over the 4 pair of
levels labeled by the $m_F$ value as defined after equation
(\ref{Aa1}) and this index is omitted everywhere. $D_0$ represents
the effect of the Zeeman phase shifts $\phi_{Z}$ on the visibility,
neglecting their spatial dispersion. $D_B$ and $N_B$  represent the
effects of the  dispersions of the diffraction phase shift $\delta
\varphi_d$ and of the Zeeman phase shift $\delta\varphi_{Z}$. The
effects of $D_0$ and $D_B$ are independent of the application of the
electric field. $D_{+/-} $ and $N_{+/-} $ are the sum of four terms
which involve the simultaneous application of the electric and
magnetic field: the first index is the parity with respect to voltage
reversal and the second index is the parity with respect to current
reversal.

In ref. \cite{LepoutrePhD11}, we have developed the calculations of
the $D_{\pm,\pm}$ and $N_{\pm,\pm}$ terms including the contributions
of the dispersion $\delta \varphi_{S,c}$  due to contact potential
(see HMWI) and the presence of this $V$-odd phase largely increases
the number of terms in these equations. As the contact potential
terms appear to be extremely small, we do not take them into account
in the present discussion but we refer the reader to ref.
\cite{LepoutrePhD11} for a more complete discussion. With this
simplification, $\delta \varphi_{S}$ is reduced to the geometrical
defect term which is $V$-even and the $D_{\pm,\pm}$ and $N_{\pm,\pm}$
terms are given by:

\begin{eqnarray}
\label{Aa6} D_{+,+}&=& \left[\frac{\left<\left(\delta
\varphi_S\right)^2\delta \varphi_Z\right>}{2}+
\left<\delta \varphi_S\delta \varphi_d \delta \varphi_Z \right>
\right]  \sin
\left( \phi_Z\right) \nonumber \\
D_{-,+} &=& D_{+,-} = 0\nonumber \\
D_{-,-} & =&  \left[ -1 + \frac{\left<\left(\delta \varphi_d  +
\delta \varphi_S \right)^2 \right>+ \left<\left(\delta
\varphi_Z\right)^2\right>}{2}\right]\nonumber \\ &\times & \varphi_{AC}
\sin \left( \phi_Z\right) + \left[\frac{\left<\left(\delta \varphi_Z
\right)^3 \right>}{6} + \left<\delta \varphi_d  \delta \varphi_S
\delta \varphi_Z \right>\right.  \nonumber \\
&+& \left.\frac{\left<\left(\delta \varphi_S\right)^2 \delta
\varphi_Z + \left(\delta \varphi_d\right)^2 \delta \varphi_Z
\right>}{2}
\right]\varphi_{AC} \cos \left( \phi_Z\right) \nonumber \\
N_{+,+} &=&  \left<\delta \varphi_S \delta \varphi_Z\right> \sin \left( \phi_Z\right) + \frac{\left<\delta \varphi_S \left(\delta \varphi_Z\right)^2\right>}{2} \cos \left( \phi_Z\right) \nonumber \\
N_{-,+} &=& N_{+,-} = 0 \nonumber \\
N_{-,-} & =&  \left[\left<\delta \varphi_S \delta \varphi_Z\right> + \left<\delta \varphi_d \delta \varphi_Z\right>\right] \varphi_{AC} \cos \left( \phi_Z\right) \nonumber \\
&-& \left[\frac{\left<\left(\delta \varphi_S +\delta \varphi_d  \right)^3 \right>}{6} + \frac{\left<\left(\delta \varphi_S +\delta \varphi_d \right)  \left( \delta \varphi_Z \right)^2 \right>}{2} \right]\nonumber \\
&&\times \varphi_{AC} \sin \left( \phi_Z\right)
\end{eqnarray}

\noindent From these equations, it is easy to deduce the relative
visibility and the phase shift of the interference fringes:

\begin{eqnarray}
\label{Aa9} \mathcal{V}_r &=& \frac{\mathcal{V}_m}{\mathcal{V}_0} =
\frac{\sqrt{D^2 + N^2}}{4} \approx \frac{D}{4} \nonumber \\
\phi_m &=&  \phi_S + \varphi_{HMW} -\theta \nonumber \\
\theta &\approx &  N/D
\end{eqnarray}

\noindent We have used a third-order approximation of the sine and
cosine function in eq. (23) of HMWI but we use only a first-order
approximation to get $\theta \approx N/D$ and $\mathcal{V}_r \approx
D/4$. This first order approximation is good if $N\ll D$. For a
practical use of these results, it will be necessary to assume that
$D_0$ is considerably larger that the other terms appearing in $D$
and that $N$ is small with respect to $D_0$ so that we will further
simplify the expression of $\theta \approx  N/D_0$. We are going to
use the following equations for the analysis of our experimental
results:

\begin{eqnarray}
\label{a2} \frac{\mathcal{V}_m(V,I)}{\mathcal{V}_0} &=& \frac{1}{4}
\left[ \left(1 -\frac{\left<\left(\delta \varphi_S + \delta \varphi_d
\right)^2\right>}{2}\right)D_0 \right.  \nonumber \\
&&  \left.  + D_Z + D_{+/-}\right] \nonumber \\
\phi_m (V,I)&=& \phi_S + \varphi_{HMW} - \frac{\left<\left(\delta \varphi_S + \delta \varphi_d \right)^3\right>}{6}\nonumber\\
&& - \frac{N_Z + N_{+/-}}{D_0}
\end{eqnarray}

\subsection{Calculation of the phase shift and the visibility neglecting the effects of the laboratory magnetic field}
\label{theory2}

To evaluate the visibility $\mathcal{V}_{EB}(V,I)$ and
the phase $\phi_{EB}(V,I)$ defined by eqs. (\ref{a1}), we
use eq. (\ref{a2}) to calculate the terms
corresponding to the different field configurations. As a first
simplified approach, we consider that the laboratory magnetic field
is homogeneous. In the case of fields configurations for which $I=0$,
apart from canceling the Zeeman phase shifts, this also enables
neglecting the effect of the Aharonov-Casher phase. At first order
in the stray terms, several cancelations appear:

\begin{eqnarray}
\label{a3}
\mathcal{V}_{EB}(V,I)&=& 1 + \frac{D_{+/-}(V,I)}{D_0(V,I)} \nonumber \\
\phi_{EB}(V,I) &=& \varphi_{HMW} (V,I) -  \frac{N_{+/-}(V,I)}{D_0(V,I)}
\end{eqnarray}

\noindent Using the definitions of eqs. (\ref{a4}), we
separate the contributions in $D_{+/-}$ and in $N_{+/-}$
following their even/odd characters with respect to $V$
and $I$. Here are the results for the visibility:

\begin{eqnarray}
\label{a5}
\mathcal{M}_E\mathcal{V}_{EB}&=& 1 +\left[\sum\left(D_{+,+} + D_{+,-}\right)/D_0\right] \nonumber \\
\Delta_E\mathcal{V}_{EB}&=& \sum\left( D_{-,+} + D_{-,-}\right)/D_0 \nonumber \\
\mathcal{M}_B\mathcal{V}_{EB}&=& 1 + \left[\sum\left( D_{+,+} + D_{-,+}\right)/D_0 \right] \nonumber \\
\Delta_B\mathcal{V}_{EB}&=&  \sum\left( D_{+,-} + D_{-,-}\right)/D_0
\end{eqnarray}

\noindent We fully separate the four $D_{\pm,\pm}$ terms by taking
means or half differences of the above quantities:

\begin{eqnarray}
\label{a6}
\mathcal{M}_B\mathcal{M}_E\mathcal{V}_{EB}&=& 1 +\left[\sum D_{+,+}/D_0\right] \nonumber \\
\mathcal{M}_B\Delta_E\mathcal{V}_{EB}&=& \sum D_{-,+}/D_0 \nonumber \\
\Delta_B\mathcal{M}_E\mathcal{V}_{EB}&=& \sum D_{+,-}/D_0 \nonumber \\
\Delta_B\Delta_E\mathcal{V}_{EB}&=& \sum D_{-,-}/D_0
\end{eqnarray}

\noindent Similar combinations with the phase $\phi_{EB}(V,I) $ also
enable the separation of the $N_{i,j}$ terms:

\begin{eqnarray}
\label{a7}
\mathcal{M}_B\mathcal{M}_E\phi_{EB}&=& - N_{+,+}/D_0 \nonumber \\
\mathcal{M}_B\Delta_E \phi_{EB}&=& - N_{-,+}/D_0 \nonumber \\
\Delta_B\mathcal{M}_E\phi_{EB}&=& - N_{+,-}/D_0 \nonumber \\
\Delta_B\Delta_E\phi_{EB}&=& \varphi_{HMW}- N_{-,-}/D_0
\end{eqnarray}

\subsection{Effect of the inhomogeneity of the laboratory magnetic field on the measurements}

We now take into account the inhomogeneity of the
laboratory magnetic field. Its main effect is to induce weak Zeeman
phase shifts, and we neglect their spatial dispersions ($\delta J_0
(y) = 0$). This brings corrections only to the $D(0,0)$ and $D(V,0)$
terms, i.e. to the visibility terms in the field configurations for
which $I=0$. It is straightforward to calculate $D(0,0)$:

\begin{eqnarray}
\label{a11}
D(0,0) &=& \left[1 -\frac{\left<\left( \delta \varphi_d \right)^2\right>}{2}\right] D_{0,B0}\nonumber \\
\mbox{with  } D_{0,B0}&=& \left[1 + \cos\left(J_0\right) + 2
\cos\left(\frac{J_0}{2}\right)\right]
\end{eqnarray}

\noindent When the electric field is applied, the residual Zeeman
phase shifts are still present ($J_0 \neq 0$). With nonzero Zeeman
phase shifts, the AC effect modifies the visibility independently of
the presence of spatial phase dispersion $\delta \varphi (y)$: this
modification is described by the leading term $-\varphi_{AC} \sin
(\phi_Z)$ in the expressions of $D_{-,-}$, eqs. (\ref{Aa6}). We thus
obtain:

\begin{eqnarray}
\label{a11BisSL}
D(V,0) &=& \left[1 -\frac{\left<\left( \delta \varphi_S +\delta \varphi_d \right)^2\right>}{2}\right] D_{0,B0} + D_{AC,B0}\nonumber \\
\mbox{with  } D_{AC,B0}&=& - \sum \varphi_{AC} \sin\left(\phi_Z\right)\nonumber \\
&=& - \varphi_{AC,B0} \left[ \sin\left(J_0\right) + \sin\left(\frac{J_0}{2}\right)\right]
\end{eqnarray}

\noindent Here $\varphi_{AC,B0}$ is the AC phase of the $F=2,m_F=2$
sub-level in the presence of the laboratory magnetic field: this
phase shift is proportional to the applied voltage $V$. In this way,
we regain the results of eq. (\ref{ex4}) for the relative
visibility $\mathcal{V}_{E}$, and we express the asymmetry $\Delta_E
\mathcal{V}_{E}$ of the visibility with voltage reversal:

\begin{eqnarray}
\label{a13}
\mathcal{V}_{E} &= & 1 -\frac{\left<\left( \delta \varphi_S \right)^2\right>}{2} - \left<\delta \varphi_S \delta \varphi_d \right> + \frac{D_{AC,B0}}{D_{0,B0}} \nonumber \\
\Delta_E \mathcal{V}_{E} &=& \frac{D_{AC,B0}(V)}{D_{0,B0}}
\end{eqnarray}

\noindent In the field configurations with $I \neq 0$, the
calculations of the type of eqs. (\ref{a2}) are not further modified.
Therefore, the presence of the laboratory magnetic field brings a
correction only to $\mathcal{V}_{EB}$ in equations (\ref{a3}), in the
following form :

\begin{eqnarray}
\label{a14}
\mathcal{V}_{EB}(V,I)&=& 1 + \frac{D_{+/-}(V,I)}{D_0(V,I)} - \frac{D_{AC,B0}(V)}{D_{0,B0}} \nonumber \\
\end{eqnarray}

%***********************************************************************

\end{document}